\documentclass[%
pra%
,reprint%
,twocolumn%
,floatfix%
,showpacs%
,amssymb, amsmath, aps]{revtex4-1}

\usepackage[latin1]{inputenc}
\usepackage{epsfig}
\usepackage{color}
\usepackage{braket}
\usepackage{dsfont}

\allowdisplaybreaks[2]

\begin{document}

\title{Temporal dynamics of stimulated emission with applications in nuclear quantum optics}

\author{Andreas Reichegger}            
\author{J\"{o}rg Evers}
\affiliation{Max-Planck-Institut f\"{u}r Kernphysik, Saupfercheckweg 1, 69117 Heidelberg, Germany}

\date{\today}


\begin{abstract}
The temporal dynamics of stimulated emission is studied, with the particular emphasis on stimulated emission induced by x-ray pulses interacting with nuclei. In typical nuclear forward scattering experiments, the short incident x-ray pulse is accompanied by a huge number of off-resonant background photons. This prompts the question, if stimulated emission can be observed in the delayed nuclear scattering signal which is emitted after the incident pulse has passed. We find that the stimulated photons essentially overlap with the stimulating pulse. To overcome this problem, we identify the reduction of the delayed scattered light intensity as alternative signature for the stimulated emission. We further study a phase-sensitive variant of stimulated emission in the low-excitation regime, which provides convenient control parameters to facilitate the detection. Finally, we analyze the possibility to observe stimulated emission in nuclei driven by free electron lasers
or synchrotron radiation sources. 
\end{abstract}

\maketitle

\section{Introduction}

Lasers~\cite{laser1,laser2} are an indispensable tool for research and engineering, as they provide a combination of unprecedented properties like spatial and temporal coherence, beam collimation over long distances, high intensities and a narrow spectrum. Thus it is not surprising that by now laser-like sources have been developed covering large parts of the electromagnetic spectrum. 

Recently, in particular x-ray and $\gamma$-ray optics has moved into the focus of research. But conventional lasers are based on stimulated emission requiring a population inversion on the lasing transition, which poses a challenge in particular at high lasing frequencies. 
This problem has been overcome to some extend by the construction of x-ray free electron lasers (FELs)~\cite{ISI:000247344500011,ISI:000281467900020,ISI:000307046800014,review-fel}, which  rely on a different operation principle: Instead of stimulated emission, light is generated from the acceleration of relativistic electron beams. Nevertheless, conventional lasers remain of interest also at x-ray and $\gamma$-ray frequencies. Interestingly, it has recently been demonstrated that FELs can in turn be used to establish a population inversion on an atomic innershell transition, which then leads to conventional amplified spontaneous emission in the x-ray domain~\cite{nina}. 

At even higher transition frequencies, a particularly challenging case is the nuclear gamma-ray laser (graser), which uses a nuclear transition as lasing transition. 
Such a device could potentially enable a large range of applications in science and technology~\cite{AnwendungKernlaser}. But dispite considerable efforts~\cite{SummaryNuklearLaser1,SummaryNuklearLaser2,ProposalNuclearLaser,BrinkeExciton}, the realization of a graser has not been achieved yet. 

Motivated by the challenges in realizing a gamma-ray laser, here, we go one step back and study the prototype process of stimulated emission itself~\cite{BuzekSE,Valente2,Shanhui,BuzekSE,Elyutin2012,Valente1}, focusing in particular on the nuclear case.
\begin{figure}[b]
 \centering
 \includegraphics[width=0.8\columnwidth]{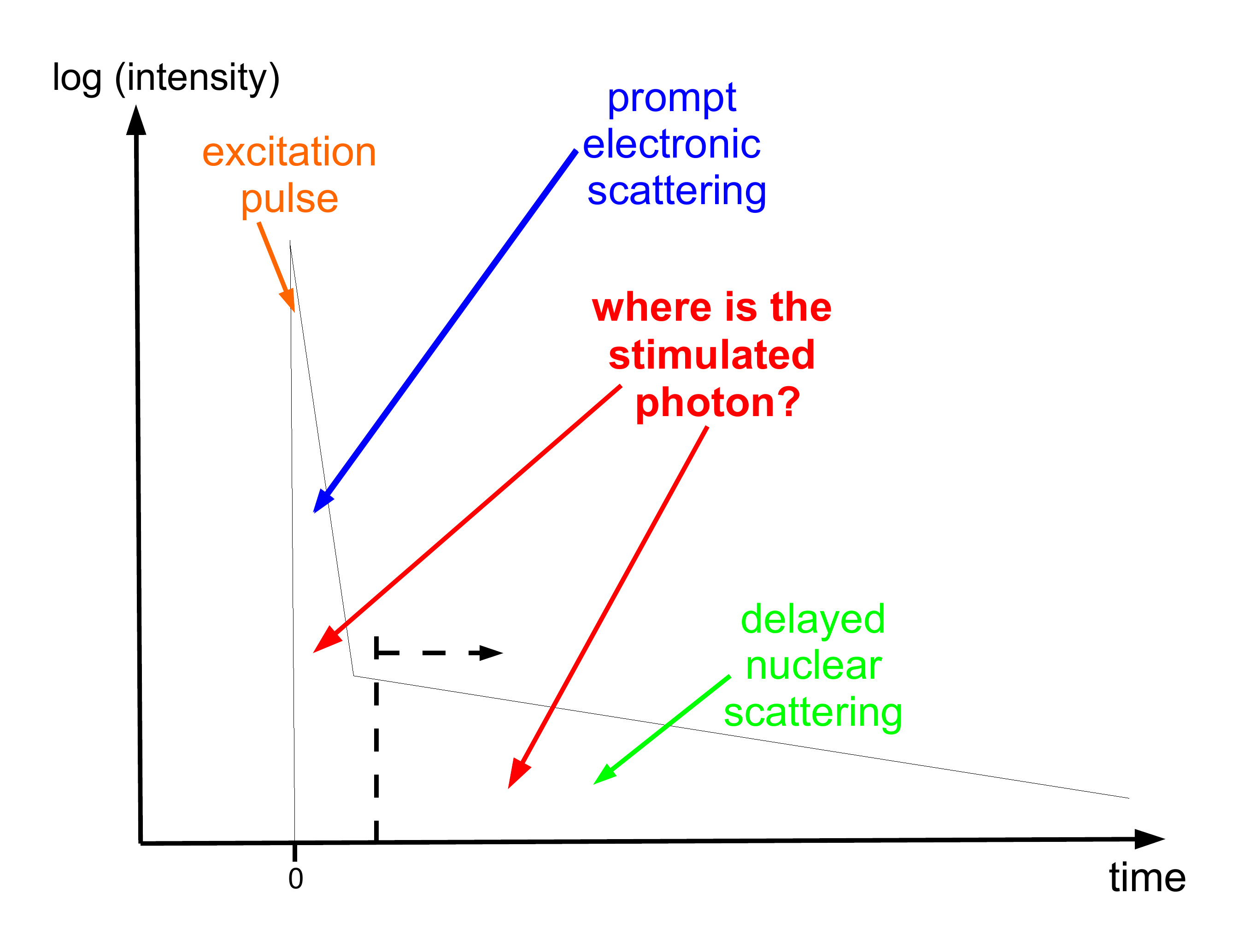}
 \caption{(Color online) Temporal evolution of nuclear forward scattering. The short incident pulse leads to an intense prompt background, which masks the nuclear signal. Removing this background by time gating (indicated by the dashed arrow which visualized the detection time interval) provides access to the pure delayed nuclear scattering signal. Motivated by the question, whether signatures of stimulated emission appear in this delayed nuclear scattering signal, here, we study the temporal dynamics of stimulated emission. }\label{Motivation}
\end{figure}
As model system, we consider nuclear forward scattering (NFS) of x-rays on large ensembles of nuclei, which has recently been demonstrated as a promising platform for nuclear quantum optics~\cite{RoehlsbergerScience,sgc,RR12,olga,PRLEvers06,Adams13,EversPRL09,PhysRevLett.109.197403,PhysRevLett.109.262502}. In synchrotron-based NFS experiments, a light pulse of $\sim$ps duration monochromatized to $\sim$meV bandwidth hits a large ensemble of nuclei. For the standard case of the ${}^{57}$Fe M\"ossbauer transition, the natural linewidth is only $4.7$neV, such that most photons of the incident synchrotron pulse form an off-resonant background. In typical experiments, this background problem is overcome by exploiting the natural nuclear life time of about 140ns which is long compared to the $\sim$ps duration of the exciting pulse, see Fig.~\ref{Motivation}~\cite{Sturhahn}. Via time gating at the detector, only the delayed photons scattered by the nuclei are recorded, which are free of the background of the incident pulse.

This immediately raises the question whether a similar time gating would be suitable for the observation of stimulated emission, and---as a consequence---what the temporal dynamics of stimulated emission is: Is the stimulated photon emitted throughout the interaction with the almost instantaneous stimulating light pulse, or is it possible to observe signatures of stimulated emission also in the delayed nuclear scattering part? 

To answer these questions, we first consider the dynamics of stimulated emission in a 1D-cavity using a numerical solution of a suitable multi-mode Jaynes-Cummings Hamiltonian. By comparing the dynamics of a freely moving photon wavepacket, of spontaneous emission, and of stimulated emission, we in detail study the temporal dynamics of stimulated emission. We find that the stimulated pulse is emitted with only a short delay throughout the interaction with the stimulating pulse. Nevertheless, we show that also the delayed part of the scattered light exhibits a distinct signature of spontaneous emission. 
In view of the expected low event rate of nuclear stimulated emission at current experimental conditions, in addition to stimulated emission from a fully inverted ensemble, we also consider stimulated emission from an ensemble of nuclei in the low-excitation regime. In contrast to the usual stimulated emission, this form of induced emission is a coherent process, which can be controlled via the relative phase of the pump and the stimulating pulse, which facilitates an experimental identification. 
Afterwards, we compare the results of the quantized theory with a corresponding semiclassical model based on the optical Bloch equations, and show that the  results of the two methods agree well. As a consequence, the analysis can be performed with the semiclassical equations, which substantially reduces the computational effort.  Based on this framework, we estimate the event rates of stimulated emission in NFS in a special thin film cavity geometry~\cite{RoehlsbergerScience} for different light sources and find that stimulted emission should be observable at a free electron laser combined with a split-and-delay unit to generate a double pulse with a fixed relative phase.

\section{Quantized description of stimulated emission}

\subsection{Model}

\subsubsection{Atom-field interaction in a cavity}

Following~\cite{BuzekKim1997,BuzekKnight,BuzekSE,Buzek2000}, we describe the interaction of a multi-mode electromagnetic field with a single two-level atom in a one-dimensional (1D) cavity using  the multi-mode Jaynes-Cummings Hamiltonian~\cite{Scully}. After transforming to the interaction picture and taking into account the dipole and rotating-wave approximation (RWA), the Hamiltonian reads 
\begin{equation}
{\hat H}^{}_{int}= \sum_{n=1}^N \hbar \Delta_{n} {\hat a}_{n}^{\dagger}{\hat a}_{n} - \sum_{n=1}^N \hbar g_{n} \left[{\hat \sigma}_{+} {\hat a}_{n} + {\hat \sigma}_{-} {\hat a}_{n}^{\dagger}\right] \ \text{.}
\end{equation}
Here, ${\hat a}_{n}$ and ${\hat a}_{n}^{\dagger}$ are the annihilation- and creation operators of the \textit{n}-th cavity mode, and ${\hat \sigma}_{-}=\ket{g}\bra{e}$ and  ${\hat \sigma}_{+}=\ket{e}\bra{g}$ are atomic lowering and raising operators between ground state $|g\rangle$ and excited state $|e\rangle$, respectively. The transition frequency of the atom is $\omega_{A}$ and $\tfrac{1}{2}\hbar\omega_{A}{\hat \sigma}_{z}$ with ${\hat \sigma}_{z}=\ket{e}\bra{e}-\ket{g}\bra{g}$ describes the spectrum of the atomic eigenstates. The zero-point energy is set to the middle of the states $\ket{e}$ and $\ket{g}$. The perfect mirrors of the cavity are placed at $z=0$ and $z=L$, where $L$ denotes the length of the cavity and therefore the momenta and frequencies of the electromagnetic field take only discrete values 
$k_{n}=\omega_{n}/c=n \pi/L$. The detuning of the \textit{n}-th cavity mode from the atomic transition frequency is
\begin{equation}
\Delta_{n}=\omega_{n}-\omega_{A}=\frac{n\pi c}{L}-\omega_{A} \ \text{.}
\end{equation}
In the model, no mechanical effects are included and therefore the atom is considered as a stationary point-like particle with an infinite mass. The coupling of the atom with the electromagnetic field is characterized by
\begin{equation}\label{Kopplung}
g_{n}=\sqrt{\frac{\omega_{n}}{\hbar\epsilon_{0}L}} \ d_{eg} \sin (k_{n}z_{A}) \ \text{,} 
\end{equation}
where $z_{A}$ is the position of the atom, which is, throughout this article, always placed in the middle of the cavity $z_{A}=L/2$. In this configuration,  only the odd modes will couple to the atom, since the coupling $g_{2n}=0$ vanishes for even modes. We rewrite the dipole matrix element $d_{eg}=\bra{e} {\hat d} \ket{g}$ in Eq.~(\ref{Kopplung}) in terms of  the one-dimensional Wigner-Weisskopf decay rate \cite{BuzekSE}
\begin{equation}\label{WWRate1D}
\Gamma_{A}=\frac{\omega_{A}|d_{eg}|^{2}}{\epsilon_{0} \hbar c} \ \text{,}
\end{equation}
to work with the more intuitive parameter $\Gamma_{A}$. The operator of the quantized electric field can be expressed as
\begin{equation}
\hat{E}(z)=\sum_{n=1}^{N}\sqrt{\frac{\hbar\omega_{n}}{\varepsilon_{0}L}} \left[\hat{a}_{n} + \hat{a}^{\dagger}_{n} \right]\sin(k_{n}z) \ \text{.}
\end{equation}

\subsubsection{Solution of the model for specific excitation subspaces}

The index $N$, which was used as an upper bound in all summations, corresponds to the use of a specific frequency cut-off \cite{BuzekKnight} required for the numerical calculations. The cutoff number $N$ depends on the choice of the parameters in the model and the fulfillment of consistency criteria. Because of the RWA, which ensures energy conservation, and $[{\hat Q}_{tot},{\hat H}]\ket{\psi}=0$, the total number of excitations 
\begin{equation}
{\hat Q}_{tot} = {\hat Q}_{A} + {\hat Q}_{F} = \ket{e}\bra{e} + \sum_{n=1}^{N} {\hat a}_{n}^{\dagger} {\hat a}_{n} \ \text{}
\end{equation}
in the atom-field system is a conserved quantity and helps us to find suitable eigenstates of the system. In the following, we will need three different wave functions for describing three physical situations with different boundary conditions. In particular, we consider the one-excitation subspace without atom, the one-excitation subspace with atom-field interaction, and the two-excitation subspace with atom-field interaction. The respective wave functions are given by
	\begin{align}
	\ket{\psi_{1} (t)}&= \sum_{n=1}^{N} A_{n}(t) \ket{1_{n}}\,. \label{WaveFunct1ExOhneInt} \\
	\ket{\psi_{2} (t)}&= B(t) \ket{e,0} + \sum_{n=1}^{N} C_{n}(t) \ket{g,1_{n}} \label{WaveFunct1Ex}\,,\\
	\ket{\psi_{3} (t)}&= \sum_{n=1}^{N} D_{n}(t) \ket{e,1_{n}} + \sum_{n=1}^{N} E_{n}(t) \ket{g,2_{n}}\nonumber \\ 
      &  + \frac{1}{2}\underset{n\neq m}{\sum_{n,m=1}^{N}} F_{nm}(t) \ket{g,1_{n},1_{m}} \ \text{.}\label{WaveFunct2Ex}
\end{align}      
Regarding, for example, the one-excitation subspace, the product eigenstate $\ket{e,0}$ describes the state, where the atom is excited and all field modes are in the vacuum state, whereas $\ket{g,1_{n}}$ describes those eigenstates where the atom is in the ground state and only the $n$-th mode is occupied with one photon and all other modes are empty.

After truncation of the Hilbert space, the system's time evolution is obtained from a direct integration of the Schr\"odinger equation in the interaction picture
\begin{equation}
i \hbar \frac{\partial}{\partial t} \ket{\psi_{i}(t)}^{} = {\hat H}_{int}^{} \ket{\psi_{i}(t)}^{} \,.
\end{equation}

\subsubsection{Observables}
From the time evolved wave functions $\ket{\psi_{i}(t)}$, we next define a number of observables which will be used throughout the later analysis. 
The probability that the upper level of the atom is excited is given by
\begin{equation}
P_{i}^{}(t) = \bra{\psi_{i}(t)} {\hat Q}_{A} \ket{\psi_{i}(t)} \ \text{.}
\end{equation}
Next, we consider the modified time-dependent decay rate of the downward transition inside the cavity. Because of the exponential character of the atom's decay it is convenient to define
\begin{equation}
\Gamma_{i}(t)=-\frac{1}{P_{i}(t)}\frac{\mathrm{d}P_{i}(t)}{\mathrm{d}t} \ \text{.}
\end{equation}
The energy density of the electromagnetic field (intensity) which allows us to analyze the important space-time propagation of radiation wave packets is
\begin{equation}\label{Intensity}
I_{i}(z,t) = \bra{\psi_{i}(t)}:\epsilon_{0} \hat{E}^{2}(z):\ket{\psi_{i}(t)} \ \text{.}
\end{equation}
Here, $:\hat{O}:$ denotes the normal ordering of a operator $\hat{O}$, applied to avoid contributions of the vacuum.
Finally, the time-dependent population of all cavity modes is denoted by 
\begin{align}
S_{i}^{}(t)= \bra{\psi_{i}(t)} \sum_{n=1}^{N} {\hat a}_{n}^{\dagger} {\hat a}_{n} \ket{\psi_{i}(t)}\,,
\end{align}
which provides the spectrum of the electromagnetic radiation inside the cavity.

\subsection{Temporal dynamics of stimulated emission}

Next, we study the temporal dynamics of stimulated emission. To isolate the effect of stimulated emission, we compare three different scenarios: The propagation of a free single-photon wave packet, the spontaneous emission of an excited atom, and the induced emission of an excited atom due to an additional photon. 

\subsubsection{Propagation of a free single-photon wave packet}\label{ChapFreeWP}

\begin{figure}[t]
 \centering
    \includegraphics[width=0.49\columnwidth]{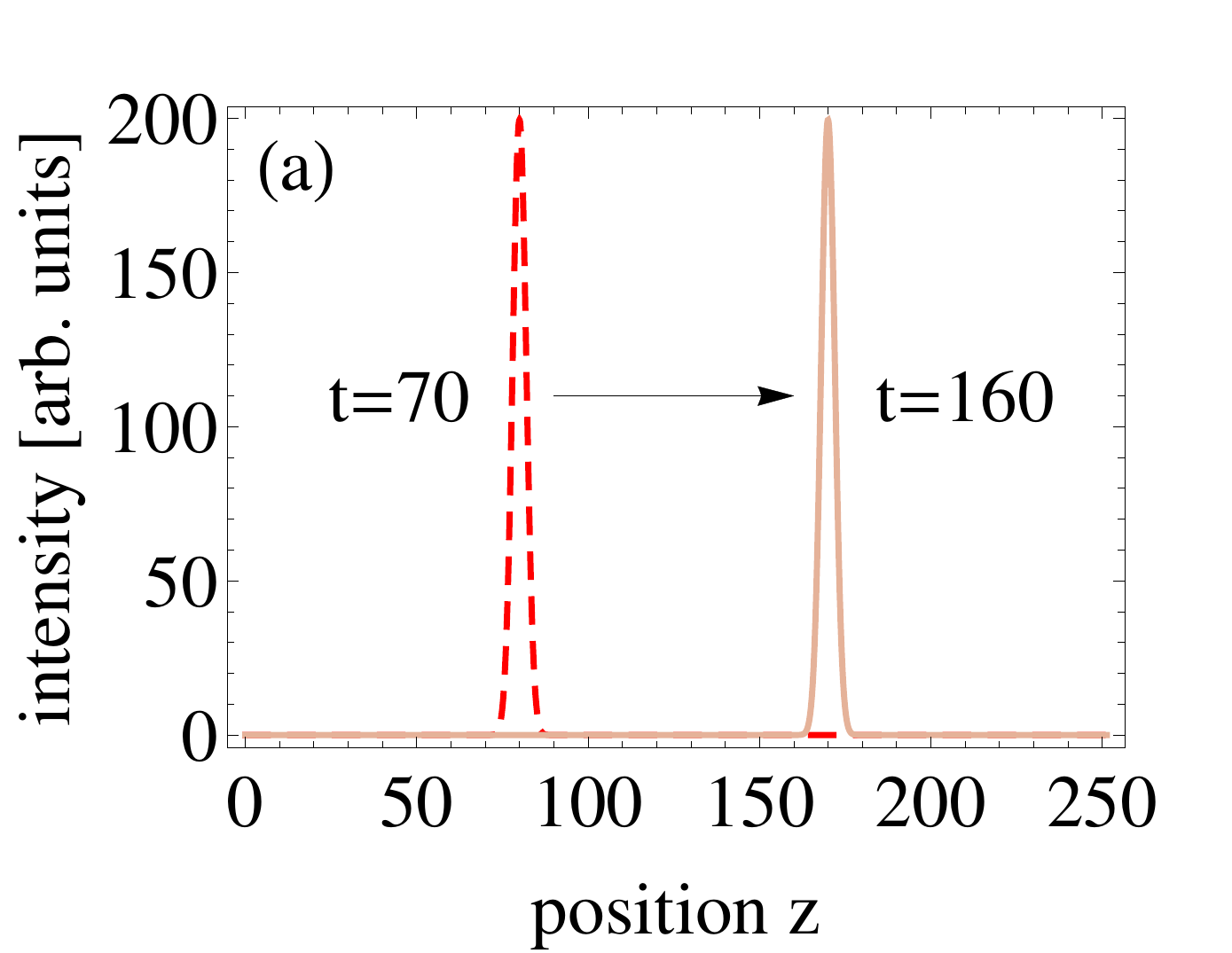}
    \includegraphics[width=0.49\columnwidth]{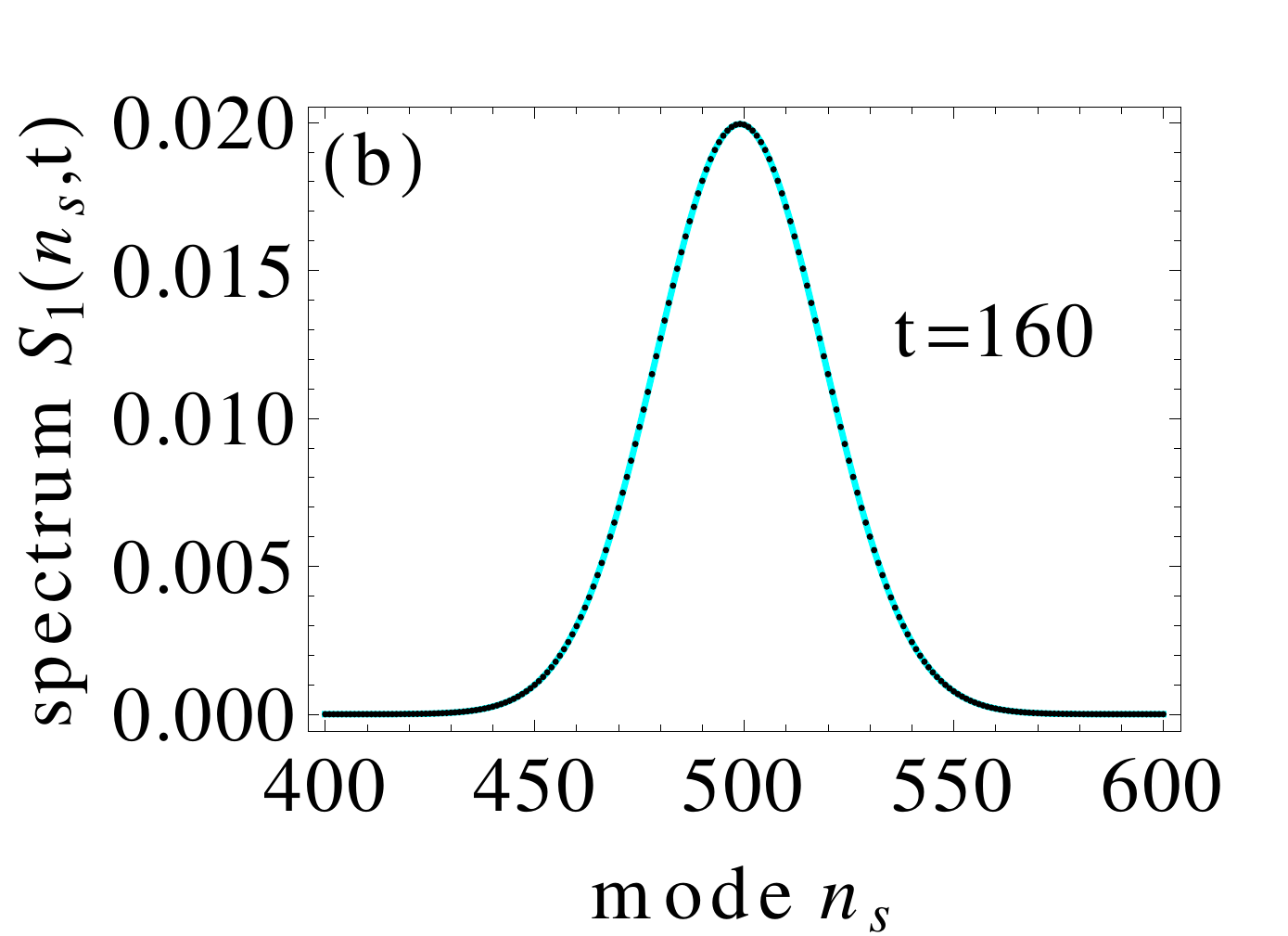}
 \caption{(Color online) (a) Intensity $I_{1}$ of a free single-photon wave packet starting its propagation at time $t=0$  from $z_{0}=10$ from left to right for two different times. (b) Cavity mode occupation in momentum space. In the numerical calculations with $N=1000$ modes, the central mode $n_s=500$ is shifted via $n_{s}=n-n_{0}+N/2$ such that it corresponds to the resonant mode $n_{0}=\omega_{0}L/(\pi c)$. The parameters in dimensionless units ($\hbar=1$, $c=1$) are $L=80\pi$, $\Gamma_{A}=0.05$, $\omega_{0}=1000$, and $\sigma=0.25$.}\label{FreeWp}
\end{figure}
The first process we study is the time evolution of a single-photon wave packet without the presence of an atom in the cavity~\cite{BuzekSE}. The time evolution of the state can be described by
\begin{equation}
\ket{\psi_{1}(t)}=U(t,0)\ket{\psi_{1}(0)}\,,
\end{equation}
where $U(t,0)=\exp(- i \hat {H}_{F}\,t/\hbar)$ is the time-evolution operator and ${\hat H}_{F}=\sum_{n=1}^N \hbar \omega_{n}{{\hat a}_{n}^{\dagger}} {\hat a}_{n}$ is the Hamiltonian of the electromagnetic field. The initial state 
\begin{equation}
\ket{\psi_{1}(t=0)}= W^{\dagger}(z_{0}) \ket{0} = \sum_{n}A_{n}(0)\ket{1_{n}}_{}
\end{equation}
is chosen in such a way that at time $t=0$ it contains a single-photon wave packet which is created at position $z=z_{0}$ by a wave packet creation operator
\begin{equation}
W^{\dagger}(z_{0}) =  \sum_{n}\frac{1}{\sqrt{\Omega_{N}}}G(k_{n})e^{-ik_{n}z_{0}}{\hat a}_{n}^{\dagger} \ \text{.}
\end{equation}
Here, we introduced a normalization factor $\Omega_{N}$ and a normalized weight function
\begin{equation}
G(k_{n}) = \frac{1}{(2\pi\sigma^{2}_{})^{1/4}} \exp{\left[-\frac{(k_{n}-k_{0})^2}{4\sigma^{2}_{}}\right]}\,,
\end{equation}
which describes a Gaussian momentum space distribution centered around $k_{0}$ with a width $\sigma$. Applying $U(t,0)$ on the initial state leads to a trivial time evolution for the amplitudes $A_{n}(t) = A_{n}(0)e^{-i\omega_{n}t}$. 
The observables are the intensity and the spectrum of the electromagnetic field, which evaluate to
\begin{eqnarray}\label{IntensFreeWP}
I_{1}(z,t) &=& \left|\sum_{n=1}^{N}M_{n}(z) A_{n}(t)\right|^{2} \,,\label{eq-i1}\\
S_{1}(n,t) &=&  \left|A_{n}(t)\right|^{2} \ \text{.}
\end{eqnarray}
Here, we introduced the position-dependent mode profile of the cavity field $M_{n}(z) =\left(\frac{2\hbar \omega_{n} }{L}\right)^{1/2} \sin(k_{n}z)$.
To ensure a suitable momentum space cutoff $N$, we verify that the conservation of the probability 
\begin{equation}\label{KonsistenzCheck}
\lim_{L \rightarrow \infty} \sum_{n=1}^{N}\left|G(k_{n})\right|^{2} \Delta k = \int_{-\infty}^{\infty} \left|G(k)\right|^{2} \mathrm dk \approx 1
\end{equation}
is fulfilled in our calculations, where the mode spacing $\Delta k= \pi/L$ . Then, the normalization factor $\Omega_{N}=L/\pi$.

In Fig.~\ref{FreeWp}~(a) the propagation of a free single-photon wave packet is illustrated. At time $t=0$ the wave packet is created at position $z_{0}=10$ and subsequently propagates from the left to the right inside the cavity. Usually in quantum  mechanics, one expects a wave packet spreading of a moving quantum object. But the two snap-shots at times $t=70$ and $t=160$ in Fig.~\ref{FreeWp}~(a) do not exhibit such a spreading. The reason for this is the 1D linear dispersion relation such that phase and group velocity are in accordance with each other. The consequence is that every mode of the wave packet will propagate in time and space with the same phase factor leading to the conservation of coherence. Note that in 2D, the dispersion relation is not linear anymore and a spreading of the wave packet appears in the direction transverse to the propagation and is suppressed in propagation direction, as discussed in \cite{Buzek2000}. Because the position space representation (intensity) shows a Gaussian shape of the pulse, we obtain the same shape in momentum space, which is depicted in Fig.~\ref{FreeWp}~(b) at time $t=160$. The Gaussian shape will not change in the course of time, because no atom is included and so the photon will not undergo any interaction during propagation. 

\subsubsection{Spontaneous decay of an excited two-level atom}

\begin{figure}[t]
 \centering
    \includegraphics[width=0.49\columnwidth]{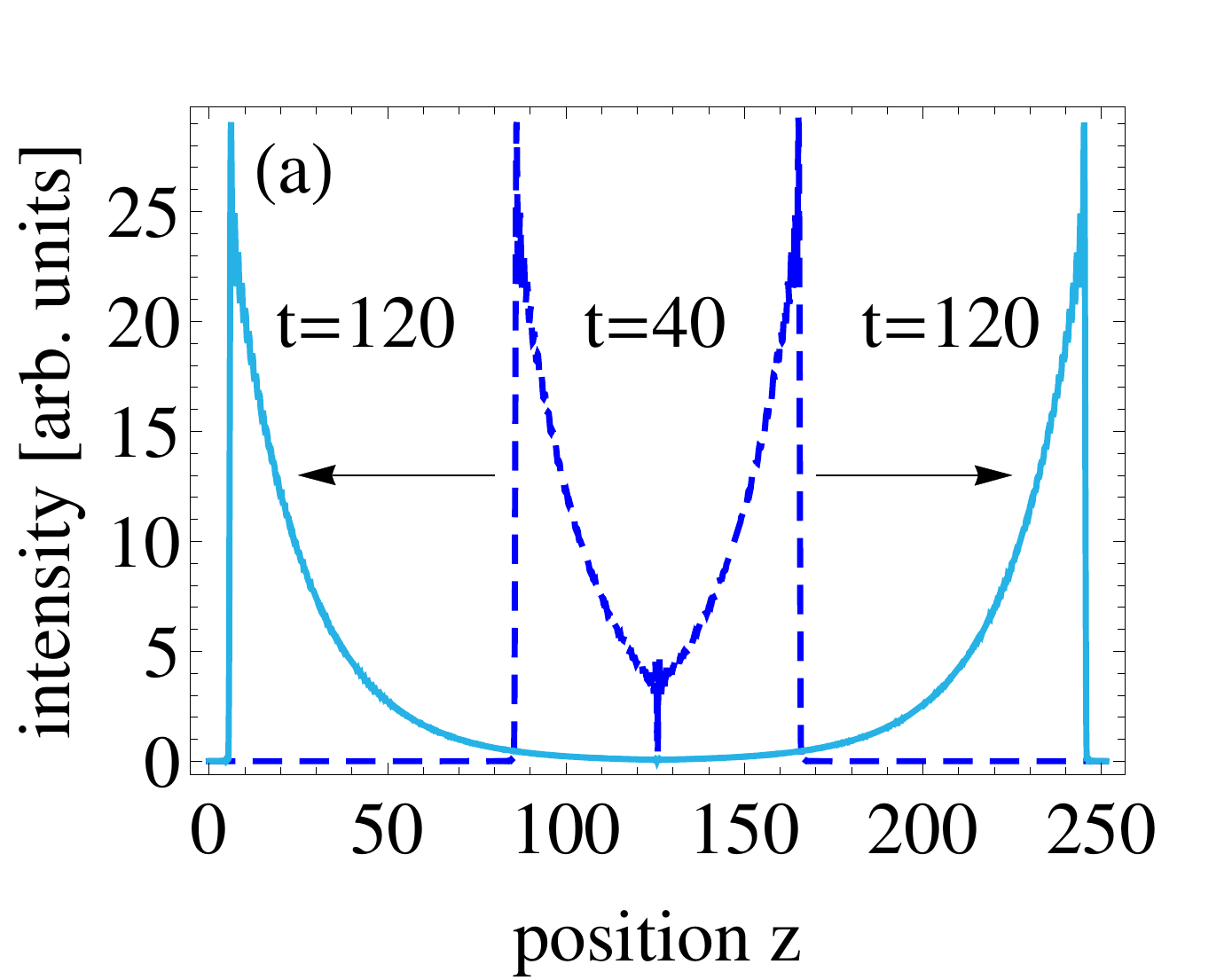}
    \includegraphics[width=0.49\columnwidth]{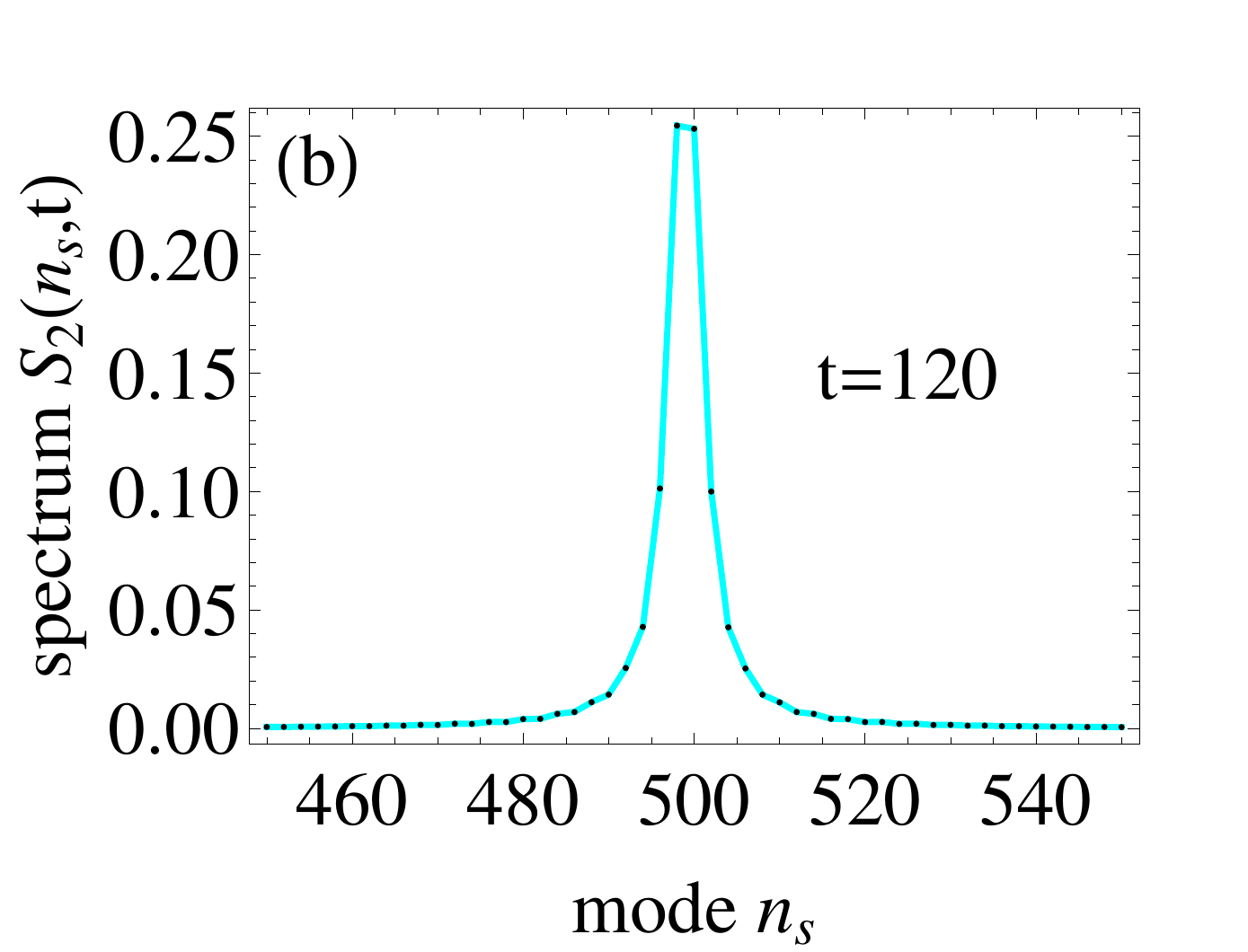}
 \caption{(Color online) (a) Intensity $I_{2}$ for spontaneous decay of an initially excited atom with wave packets propagating towards the cavity mirrors for two different times. (b) Cavity mode occupation in momentum space after the atom has radiated all of its excitation into the cavity at time $t=160$. $n_s$ is defined as in Fig.~\ref{FreeWp}. The parameters in dimensionless units ($\hbar=1$, $c=1$) are $L=80\pi$, $\Gamma_{A}=0.05$, $\omega_{A}=1000$, $\sigma=0.25$ and $N=1000$.}
 \label{SponDecay}
\end{figure}

Next, we consider spontaneous decay of an atom in the cavity, which in general will be different from free space due to the modified photonic density of states. From the Schr\"odinger equation, we obtain a set of coupled linear differential equations for the state amplitudes
\begin{subequations}
\begin{eqnarray}
i\frac{\mathrm{d}}{\mathrm{d}t}B(t)&=&-\sum_{n=1}^{N}g_{n}C_{n}(t)\label{DGLSponDecay1}  \,,\\
i\frac{\mathrm{d}}{\mathrm{d}t}C_{n}(t)&=&\Delta_{n}C_{n}(t)-g_{n}B(t)\label{DGLSponDecay2} \,.
\end{eqnarray}
\end{subequations}
For our analysis of spontaneous decay we start with an excited atom and the cavity modes in the vacuum state, $B(0)=1$ and $C_{n}(0)=0$.
The intensity, spectrum and atom excitation probability evaluate to
\begin{subequations}
\begin{eqnarray}\label{IntensSD}
I_{2}(z,t) &=& \left|\sum_{n=1}^{N}M_{n}(z) C_{n}(t)\right|^{2} \,, \label{eq-i2}\\ 
S_{2}(t) &=& \left\{ \left|C_{n}(t)\right|^{2} \right\} \,,\\
P_{2}(t)&=&|B(t)|^{2} \ \text{.} \label{eq-p2}
\end{eqnarray}
\end{subequations}
An example for the resulting time evolution is shown in Fig.~\ref{SponDecay}(a). The spontaneous decay process is accompanied by the symmetric emission of two wave packets propagating to the left and right mirror of the cavity. As expected, they consist of sharp propagating intensity fronts with an exponentially decreasing tail, corresponding to the exponential decay of the atom. The snapshot at time $t=40$ shows the atom throughout its de-excitation, whereas at $t=120$ all the excitation is already emitted to the cavity field. The sharpness of the fronts is limited by the finite number of modes included in the calculation.  After the atom has radiated all of its excitation to the cavity modes, the spectrum of the electromagnetic field shows a Lorentz distribution as sketched in Fig.~\ref{SponDecay}(b). Note that in contrast to free space, after a reflection of both wave packets at the cavity mirrors, they will be partially reabsorbed by the atom throughout the further evolution~\cite{BuzekKnight}.

\subsubsection{Interaction of a single-photon wave packet with an initially excited two-level atom}\label{ChapIntSE}

To model stimulated emission we have to expand our calculation to the two-excitation subspace with wave function $\ket{\psi_{3}(t)}$ in Eq.~(\ref{WaveFunct2Ex}). The resulting equations of motion are 
\begin{subequations}
\begin{eqnarray}
i\frac{\mathrm{d}}{\mathrm{d}t}D_{n}(t)&=& \Delta_{n}D_{n}(t) -\underset{(n \neq m)}{\sum_{m=1}^{N}} g_{n} F_{nm}(t)\nonumber \\  & &-\sqrt{2}g_{n}E_{n}(t) \,,   \\
i\frac{\mathrm{d}}{\mathrm{d}t}E_{n}(t)&=&  2\Delta_{n}E_{n}(t)-\sqrt{2}g_{n}D_{n}(t) \,,\\ 
i\frac{\mathrm{d}}{\mathrm{d}t}F_{nm}(t)&=& \left[\Delta_{n}+\Delta_{m}\right]F_{nm}(t)-g_{n}D_{m}(t) \nonumber \\ & & -g_{m}D_{n}(t) \ \text{.}
\end{eqnarray}
\end{subequations}
The initial conditions for stimulated emission require one excitation in form of a Gaussian pulse in the field and an initially excited atom:
\begin{subequations}
\begin{eqnarray}
D_{n}(0)&=& \frac{1}{\sqrt{\Omega_{N}}}e^{-ik_{n}z_{0}}G(k_{n}) \,, \\
E_{n}(0)&=&  0 \,,\\
F_{nm}(0)&=& 0 \ \text{.}  
\end{eqnarray}
\end{subequations}
We assume resonant interaction  $\omega_{0}=\omega_{A}$, which means that the central frequency of the wave packet matches the transition frequency of the atom. The observables evaluate to
\begin{subequations}
\begin{eqnarray}\label{IntensSE}
I_{3}(z,t) &=& \left| \sum_{n=1}^{N}M_{n}(z)D_{n}(t)\right|^{2} + 2 \sum_{n=1}^{N} \left|M_{n}(z)E_{n}(t)\right|^{2}\nonumber \\
& & + \sqrt{2} \underset{(n \neq m)}{\sum_{n,m=1}^{N}} M_{n}(z)M_{m}^{*}(z)  \nonumber \\ & & \ \ \ \Big[ E_{m}^{*}(t)F_{mn}(t)+ E_{n}(t)F_{mn}^{*}(t) \Big] \nonumber \\
& & + \underset{(k \neq m,n)}{\sum_{k,n,m=1}^{N}} M_{n}(z)M_{m}^{*}(z)F_{kn}(t)F_{mk}^{*}(t) \,,\label{eq-i3}\\
S_{3}(t)&=&    \left| D_{n}(t) \right|^{2} + \left| E_{n}(t) \right|^{2} + \sum_{m=1}^{N} \left| F_{nm}(t) \right|^{2}   \,,\\
P_{3}(t)&=&\sum_{n=1}^{N}\left|D_{n}(t)\right|^{2} \ \text{.}
\end{eqnarray}
\end{subequations}

\begin{figure}[t]
 \centering
    \includegraphics[width=0.49\columnwidth]{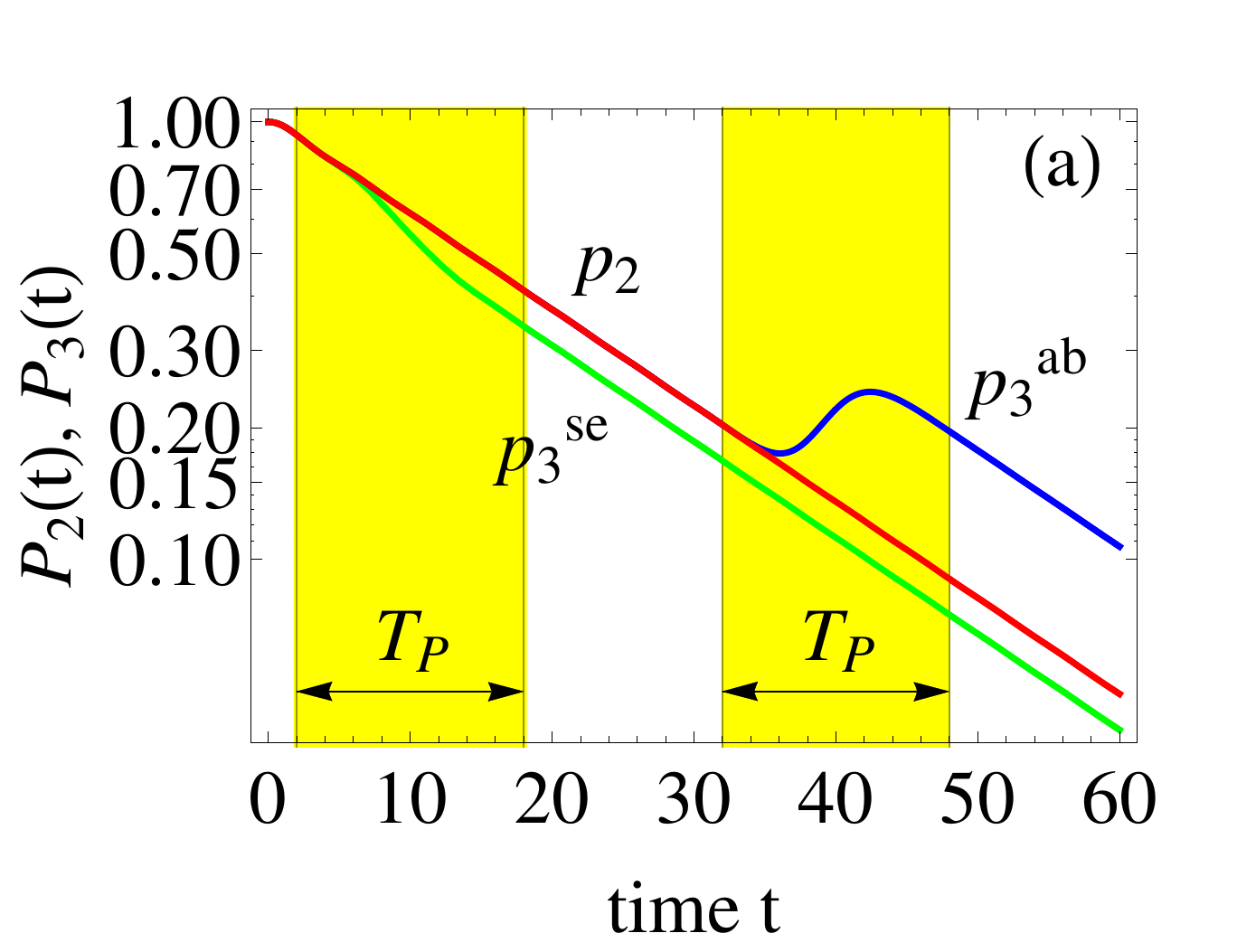}
    \includegraphics[width=0.49\columnwidth]{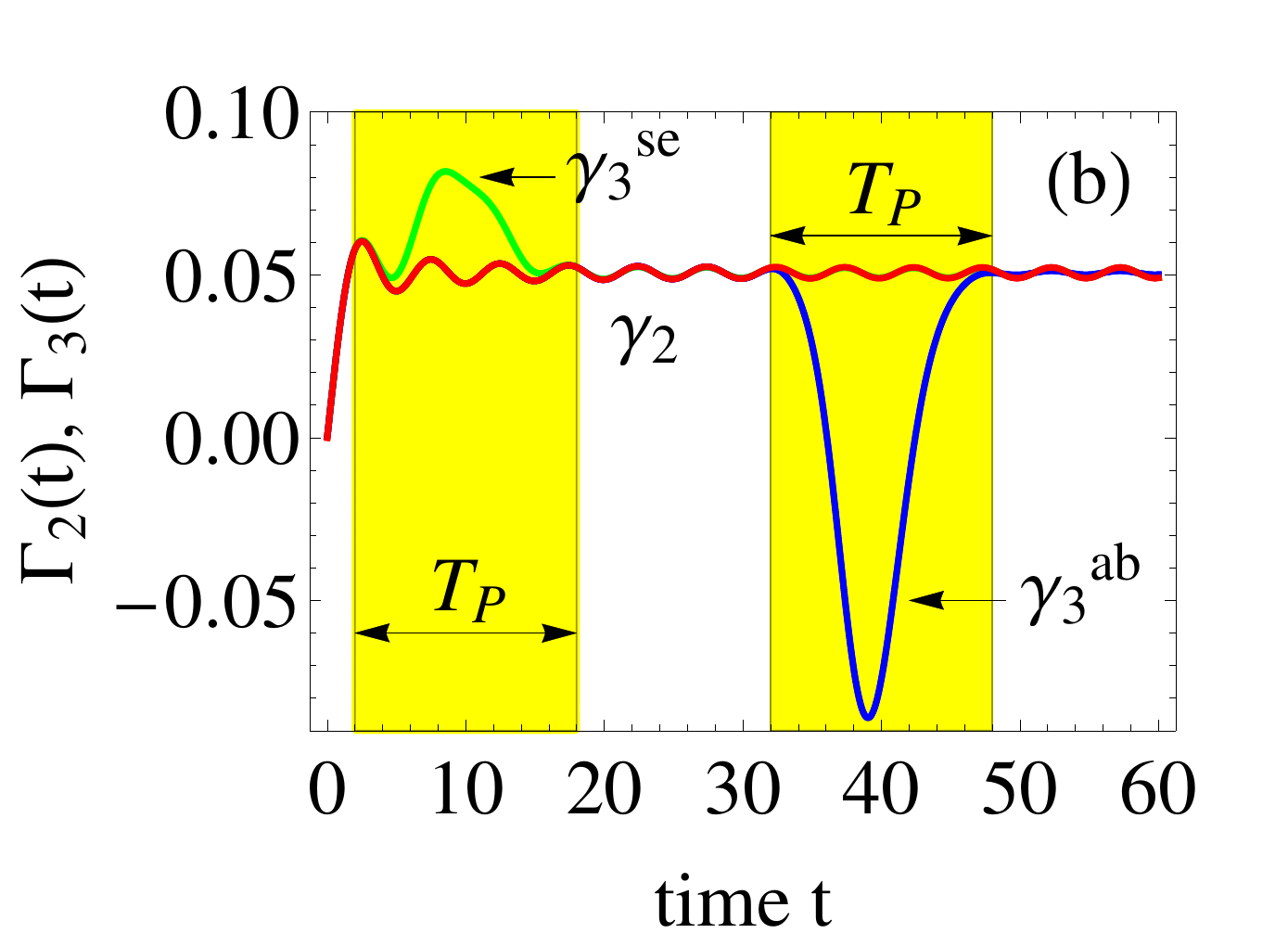}
 \caption{(Color online) (a) Time evolutions of the population of the upper atomic level $P_{i}(t)$. $p_{2}$ shows the pure exponential decay $P_{2}(t)$ of the unperturbed atom, whereas the graphs $p_{3}^{se}$ and $p_{3}^{ab}$ show the modified decay behavior $P_{3}(t)$ due to the interaction with a single-photon pulse starting its propagation at $z_{0}=117.7$ or $z_{0}=87.7$, respectively. (b) Time dependent decay rates for the results in (a). The line $\gamma_{2}$, representing the rate $\Gamma_{2}(t)$ from the pure spontaneous decay is nearly constant throughout the time evolution, whereas the graphs $\gamma_{3}^{se}$ and $\gamma_{3}^{ab}$ show a modified behavior due to the interplay with the single-photon pulse.  The colored areas mark the interaction time intervals, given by the pulse duration $T_{P}\approx16$. The parameters in dimensionless units ($\hbar=1$, $c=1$) are $L=80\pi$, $\Gamma_{A}=0.05$,  $\omega_{0}=\omega_{A}=1000$, $\sigma=0.25$ and $N=200$ modes.}
 \label{BeschStimEmiss}
\end{figure}

To analyze the effect of the additional photon on the atom dynamics, we next compare three different cases. First, the exponential atomic decay without additional photon given by $P_{2}(t)$ defined in Eq.~(\ref{eq-p2}). Second, the atomic evolution  $P_{3}(t)$ with a photon arriving early in the otherwise exponential decay of the atom, with the photon pulse initially positioned directly in front of the atom at $z_{0}=L/2-T_{P}/2=117.7$. Here, $T_{P}\approx4/(c\sigma)$ is the pulse duration. Third, the evolution  $P_{3}(t)$ with a photon pulse arriving late in the otherwise exponential decay of the atom, with the photon pulse initially positioned directly in front of the atom at $z_{0}=87.7$. The three results denoted as $p_2$, $p_{3}^{se}$ and $p_{3}^{ab}$, respectively, are shown in Fig.~\ref{BeschStimEmiss}~(a). 
It can be seen that throughout the interaction of the additional photon with the atom, the atomic population can either decrease ($p_{3}^{se}$) or increase ($p_{3}^{ab}$). After the interaction with the pulse, the atom continues with its exponential decay.

The observed shortening of the atomic lifetime in graph $p_{3}^{se}$ caused by the single-photon pulse is an indicator of stimulated emission. The re-population in $p_{3}^{ab}$, however, corresponds to net absorption by the atom. This is possible for pulses arriving at a later time, because the atom is not inverted any more: it lost most of its excitation due to spontaneous emission and therefore it can absorb radiation again. 

To further interpret these results, we also analyze the time dependent decay rate in Fig.~\ref{BeschStimEmiss}~(b). The graph $\gamma_{2}$ represents the time-dependent decay rate $\Gamma_{2}(t)$ of the unperturbed atom undergoing a pure spontaneous decay. As soon as the excited atom starts radiating its excitation to the cavity field, the time-dependent decay rate starts to grow, beginning at $\Gamma_{2}(t=0)=0$ and it reaches the value given by Fermi's golden rule after a short period of time. The oscillations of the function $\Gamma_{2}(t)$ are a numerical artifact due to the finite number of discrete modes, which will decrease for an increasing of the mode number. The line $\gamma_{3}^{se}$ corresponds to the calculated rate $\Gamma_{3}(t)$ with the initial condition that the pulse is directly positioned in front of the atom at $t=0$. During the interaction with the single-photon pulse the maximum value of $\gamma_{3}^{se}$ exceeds the rate known from Fermi's golden rule by a factor of $1.63$. This is in accordance to the mentioned acceleration of the downward transition in Fig.~\ref{BeschStimEmiss}~(a) and also indicates stimulated emission, because the interaction with the stimulating pulse leads to a shortening of the atomic lifetime. The graph $\gamma_{3}^{ab}$ belongs to the function $\Gamma_{3}(t)$ calculated for the more distant initial photon pulse and depicts another behavior. Here, the decay rate $\Gamma_{3}(t)$ first decreases, which demonstrates the suppression of radiation. But after some time, the graph $\gamma_{3}^{ab}$ even takes negative values which corresponds to the absorption of radiation from the cavity field. Now, in contrast to the described case of stimulated emission, the single-photon pulse acts as an excitation pulse which extends the atomic lifetime.

\subsubsection{Time resolved analysis of stimulated emission}

\begin{figure}[t]
 \centering
    \includegraphics[width=0.49\columnwidth]{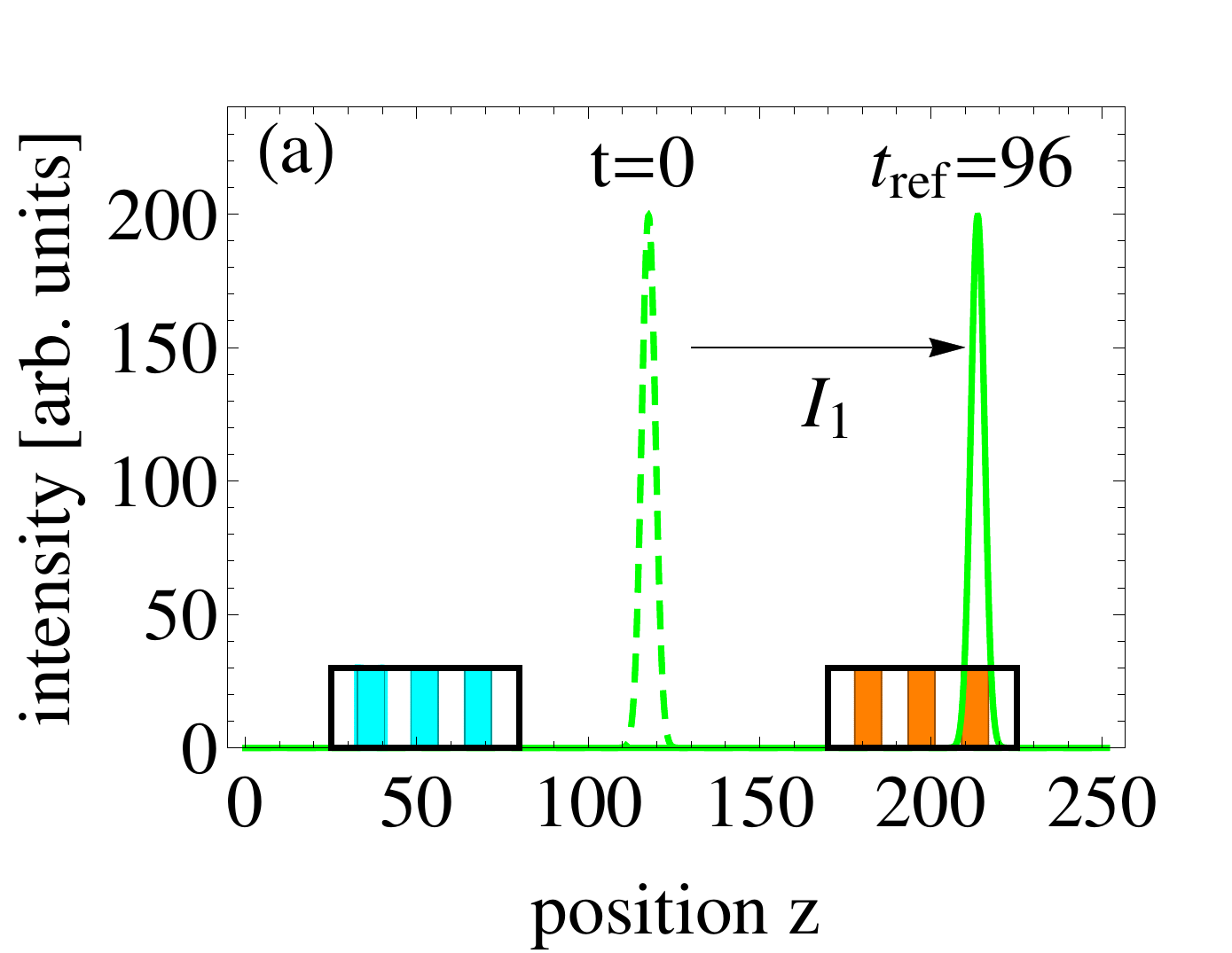}
    \includegraphics[width=0.49\columnwidth]{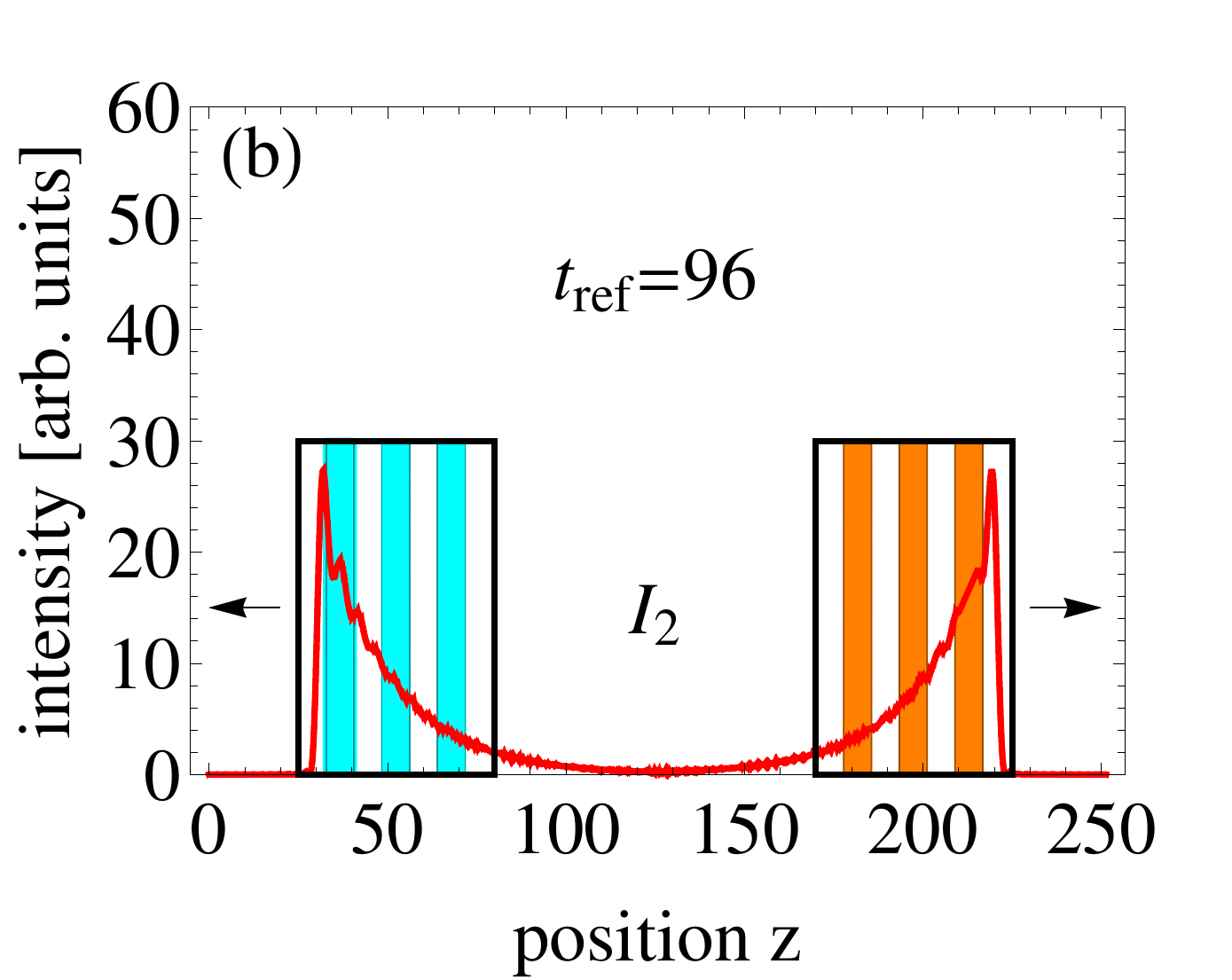}
    \includegraphics[width=0.49\columnwidth]{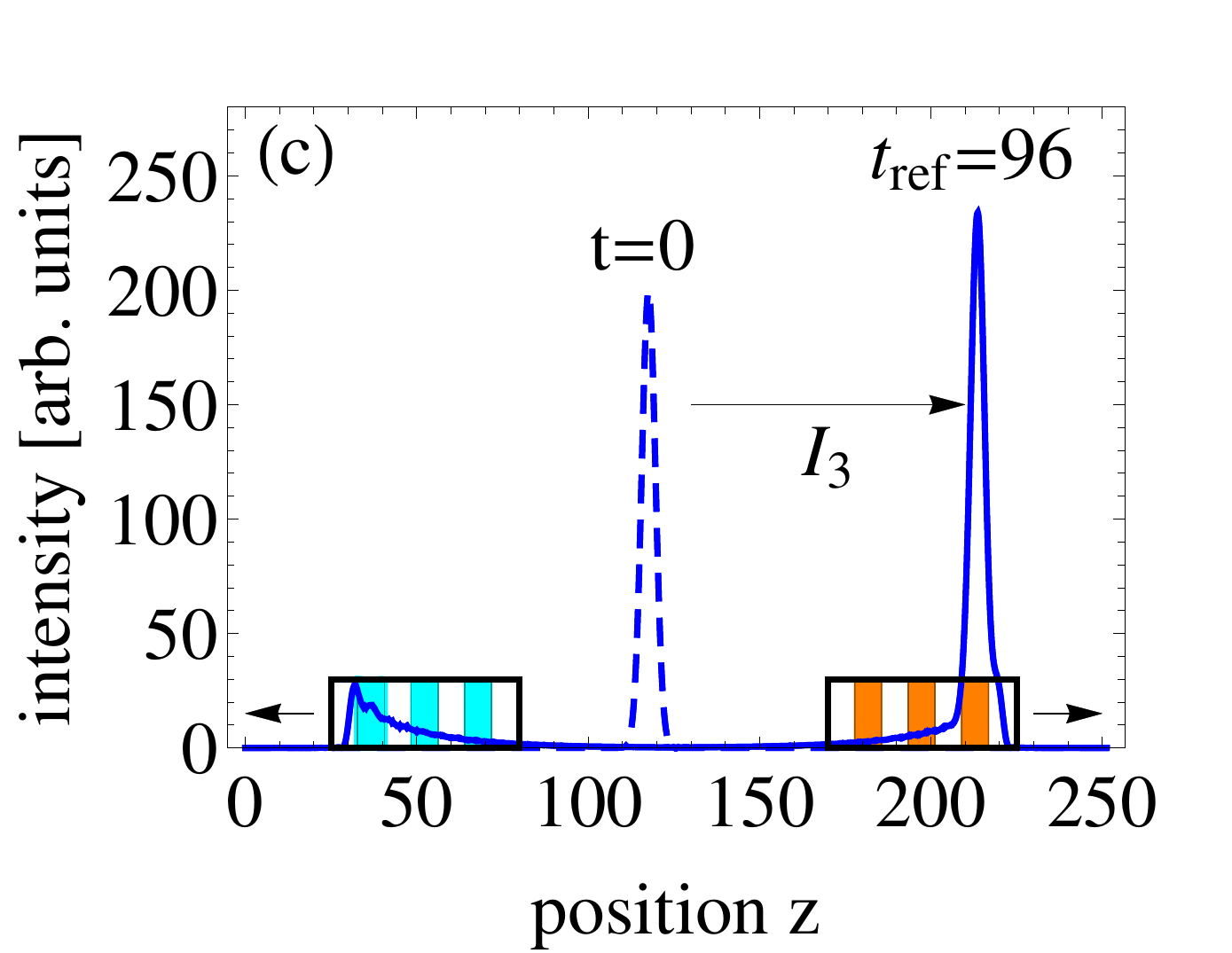}
    \includegraphics[width=0.49\columnwidth]{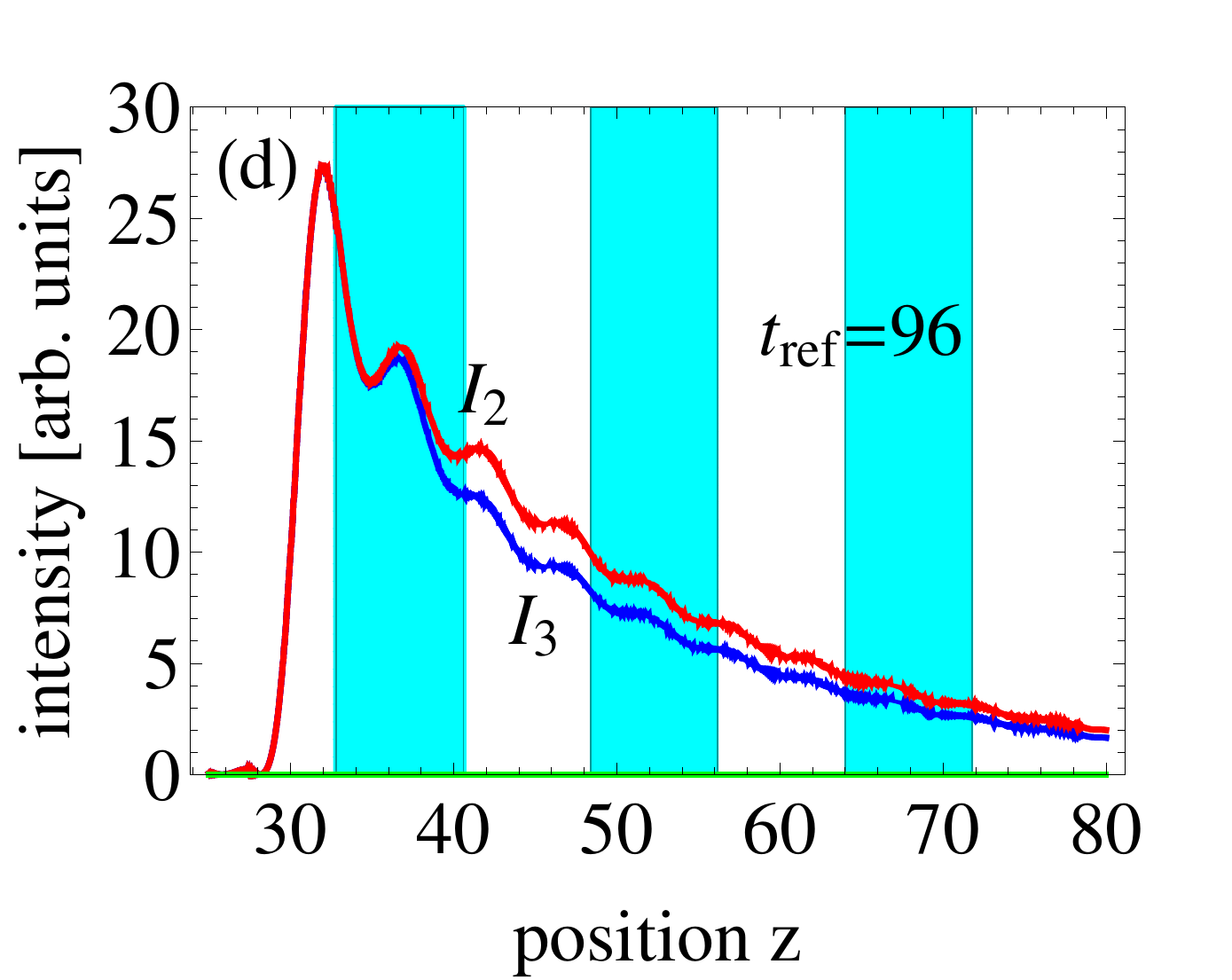}
    \includegraphics[width=0.49\columnwidth]{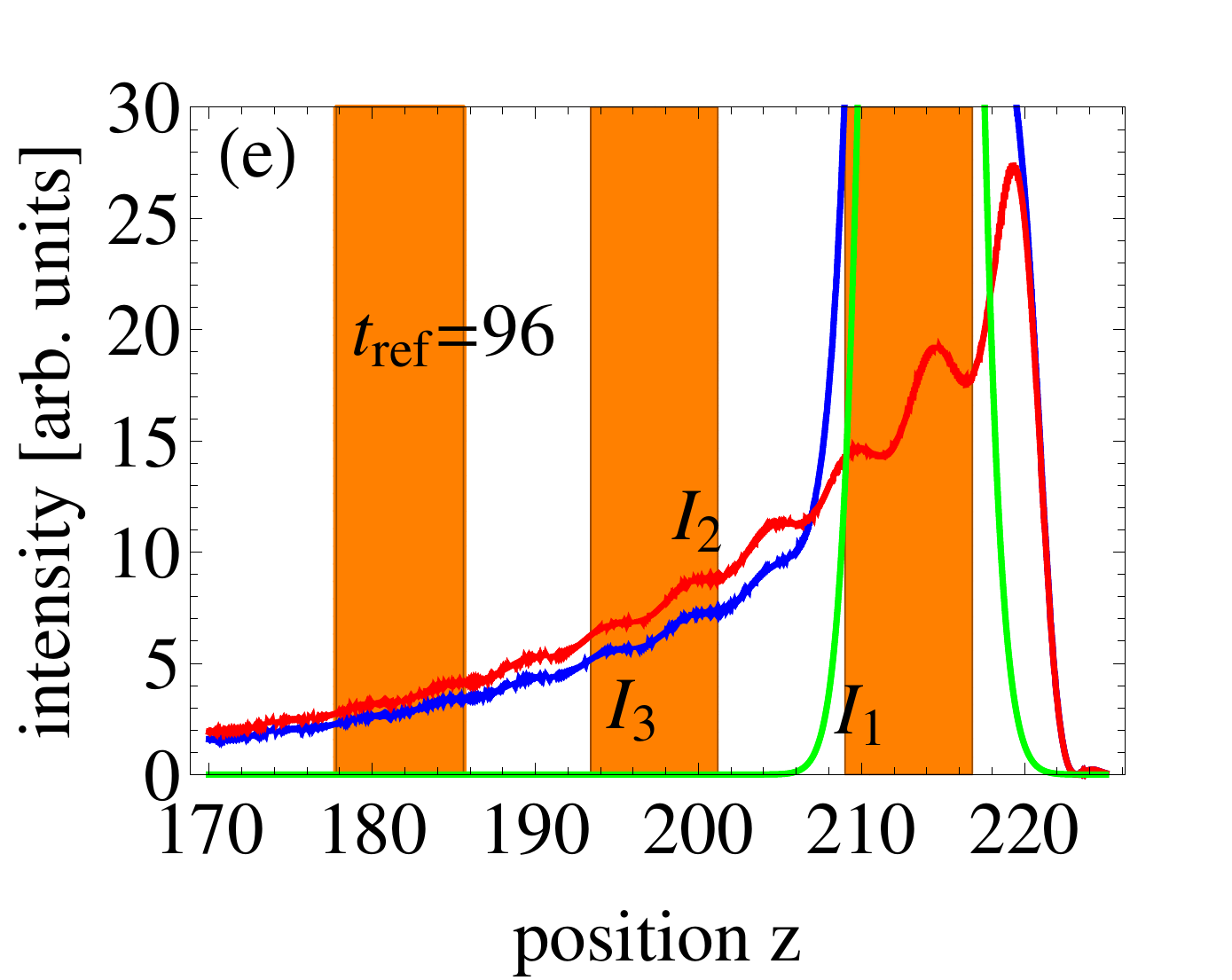}
    \includegraphics[width=0.49\columnwidth]{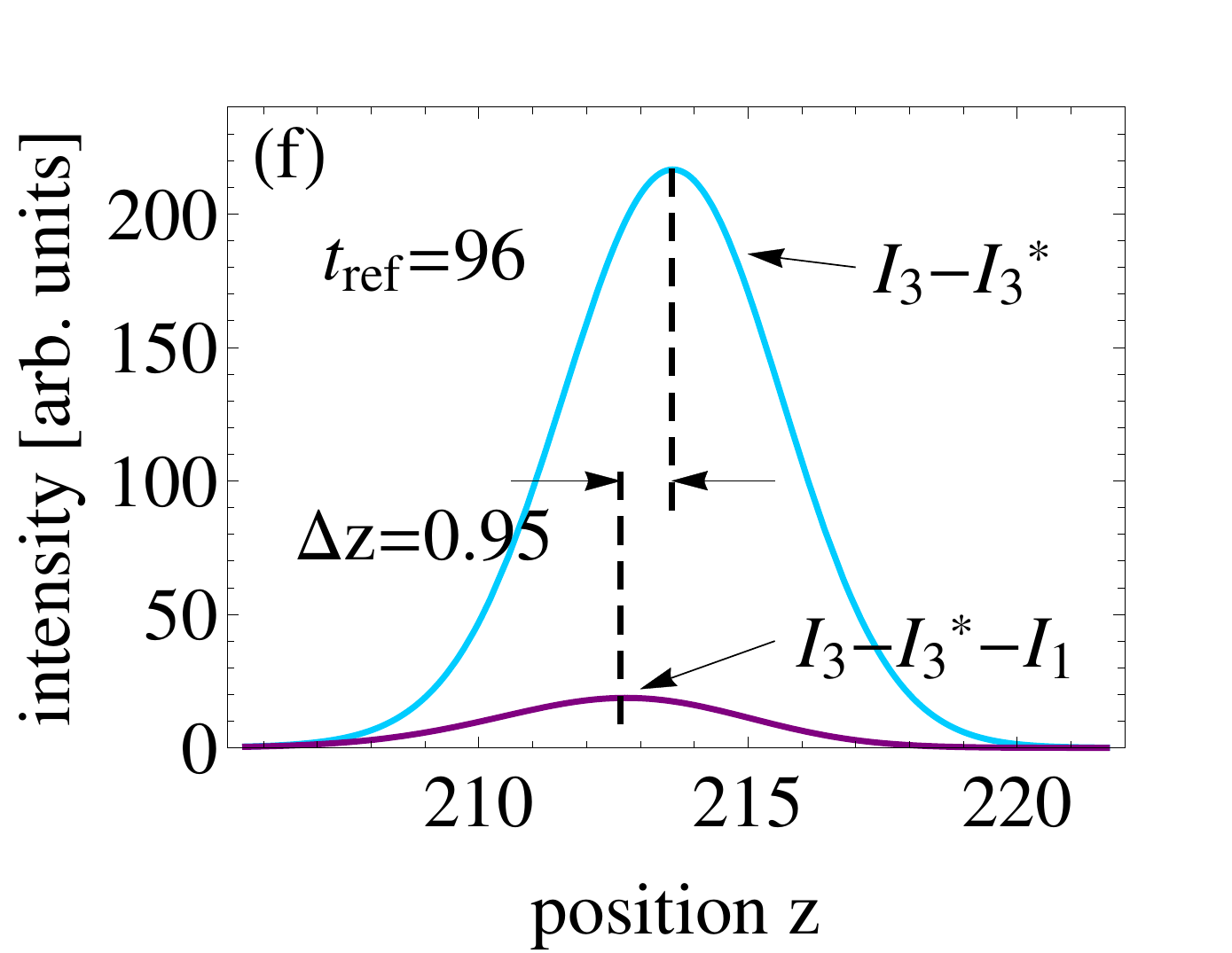}
 \caption{(Color online) (a) Intensity $I_1$ of the free wave packet propagation at times $t=0$ (dashed line) and $t_{ref}=96$ (solid line). (b) Spontaneous emission intensity $I_{2}$ at time $t_{ref}=96$. (c) The graph $I_{3}$ shows the intensity of a single-photon wave packet before ($t=0$, dashed line) and after ($t_{red}=0$, solid line) interacting with an initially excited atom. (d) and (e) show $I_{1}$, $I_{2}$ and $I_{3}$ in enlargements of the shaded areas in (a)-(c) at $t_{ref}=96$. (f) The graph of the stimulating pulse after subtraction of the modified spontaneous decay background $I_{3}-I^{*}_{3}$ and the stimulated photon pulse after subtraction of the respective backgrounds $I_{3}-I^{*}_{3}-I_{1}$. The parameters in dimensionless units ($\hbar=1$, $c=1$) for all six plots are $z_{0}=117.7$, $L=80\pi$, $\Gamma_{A}=0.05$, $\omega_{0}=\omega_{A}=1000$, $\sigma=0.25$ and $N=200$ modes.}
 \label{TimeResStimEmiss}
\end{figure}

To investigate the effect of stimulated emission further, we next analyze the spatio-temporal dynamics of the electromagnetic field. To isolate the effect of stimulated emission, we now compare the free propagation of a single-photon wave packet, the spontaneous decay of an unperturbed atom, and an initially excited atom interacting with a single-photon pulse. An example at a certain reference time $t_{ref}$ after the interaction with the stimulating pulse is shown in Fig.~\ref{TimeResStimEmiss}. 
Panel~(a) shows the intensity $I_{1}$ [Eq.~(\ref{eq-i1})] of the single-photon wave packet propagating from  the left to the right without interacting with an atom as already described in the previous section \ref{ChapFreeWP}. The snapshot of the intensity $I_{2}$ [Eq.~(\ref{eq-i2})] at time $t_{ref}=96$ in Fig.~\ref{TimeResStimEmiss}~(b) represents the pure spontaneous decay of an excited atom. The intensity $I_{3}$ [Eq.~(\ref{eq-i3})]  in Fig.~\ref{TimeResStimEmiss}(c) includes all considered processes: the interaction of an excited atom with a single-photon pulse a well as the simultaneously ongoing process of spontaneous emission of an excited atom. The dashed line in Fig.~\ref{TimeResStimEmiss}(c) shows the single-photon pulse placed directly in front of the position of the atom at $t=0$. At time $t_{ref}$, there is an enhancement of the single-photon pulse in the right half of the cavity, which is an indicator for stimulated emission. This enhancement is a direct consequence of the effect the pulse performs on the atom: it forces the excited atom to emit radiation in the direction of the propagating pulse.

To quantify the stimulated emission, we next compare the regular spontaneous emission to the induced emission case. To this end, Fig.~\ref{TimeResStimEmiss}(d) and (e)  display an enlargement of the shaded boxes seen in Fig.~\ref{TimeResStimEmiss}(a)-(c) showing the intensities $I_{1}$, $I_{2}$ and $I_{3}$ in the same plot. From this comparison, different signatures of stimulated emission can be identified. The first is the already mentioned enhancement of the intensity in the leading edge of the pulse in the right hand side of the cavity. But in addition, it can be seen that outside this leading pulse, the intensity of the modified spontaneous decay $I_{3}$ is reduced compared to the pure spontaneous decay $I_{2}$. This reduction also appears in the left hand side of the cavity. It is due to the fact that the stimulating pulse forces the atom to radiate a part of its excitation into the propagation direction during the interaction, such that there is less atomic excitation left which can be spontaneously emitted afterwards. 

This second signature is of particular relevance for the case of a broadband excitation pulse such as in nuclear forward scattering, since then the first signature of an enhanced leading pulse front may be masked by a large number of non-resonant background photons which cannot be discriminated by the detection. 
The reduction further enables us to determine the amount of induced emission. For this we calculate the intensity differences in both halves of the cavity after the interaction has taken place at a certain reference time $t_{ref}$. The intensity difference in the left half $\Delta I_{left}$ is calculated by subtraction the reference intensities $I_{1} + I_{2}$ from the full intensity $I_{3}$ at the  reference time,
\begin{equation}
\Delta I_{left} = \int_{0}^{\frac{L}{2}}  I_{3}(z,t_{ref})- [ I_{2}(z,t_{ref}) + I_{1}(z,t_{ref}) ] \:  \mathrm{d}z \ \text{.}
\end{equation}
This quantity is negative ($\Delta I_{left}<0$), because the intensity background of the free single-photon wave packet
$I_{1}$ is zero in the cavity's left half, and the pure spontaneous decay case $I_{2}$ exceeds $I_{3}$  as discussed. For the calculation of the intensity difference in the right half of the cavity, one finds analogously
\begin{equation}
\Delta I_{right} = \int_{\frac{L}{2}}^{L}  I_{3}(z,t_{ref}) - [ I_{2}(z,t_{ref}) + I_{1}(z,t_{ref}) ] \:  \mathrm{d}z \ \text{.}
\end{equation}
This quantity is positive ($\Delta I_{right}>0$), because the single-photon wave packet is propagating towards the right half of the cavity and the stimulating pulse has forced the atom to emit radiation into the propagation direction. The absolute value of the intensity difference in total $\Delta I_{tot}$ is a measure for the intensity of the induced single-photon wave packet
\begin{equation}
\Delta I_{tot} = \left|\Delta I_{left}\right| + \Delta I_{right} \ \text{,}
\end{equation}
because $\Delta I_{tot}$ describes the changes in the intensity profile in the left and right half of the cavity due to stimulated emission and thus is directly connected to the intensity of the stimulated photon. To visualize this induced single-photon wave packet in space at the mentioned reference time, we need to subtract the intensity background of the free single-photon wave packet $I_{1}$ and the intensity background of the modified spontaneous decay from the intensity $I_{3}$. The modification of spontaneous emission is not directly accessible in the right hand side of the cavity, since it occurs in parallel with the induced emission. But exploiting the mirror symmetry for spontaneous emission of an atom placed in the middle of the cavity, we can evaluate the modified emission from the left side of the cavity. For this we define the intensity $I^{*}_{3}$ which is the modified intensity of the left side of the cavity mirrored to the right side. The induced emission wave packet is then given by $I_{3}(z,t_{ref}) - I^{*}_{3}(z,t_{ref})-I_{1}(z,t_{ref})$. The result is shown in Fig.~\ref{TimeResStimEmiss}~(e), where we see the stimulating pulse after subtraction of the modified spontaneous decay background $I_{3}-I^{*}_{3}$ and the induced single-photon wave packet $I_{3}-I^{*}_{3}-I_{1}$ at the reference time $t_{ref}$. We find that the up to a small relative displacement, the induced wavepacket essentially overlaps with the stimulating pulse.

\subsection{Interaction of two single-photon wave packets with an atom in the ground state}

\begin{figure}[t]
    \includegraphics[width=0.49\columnwidth]{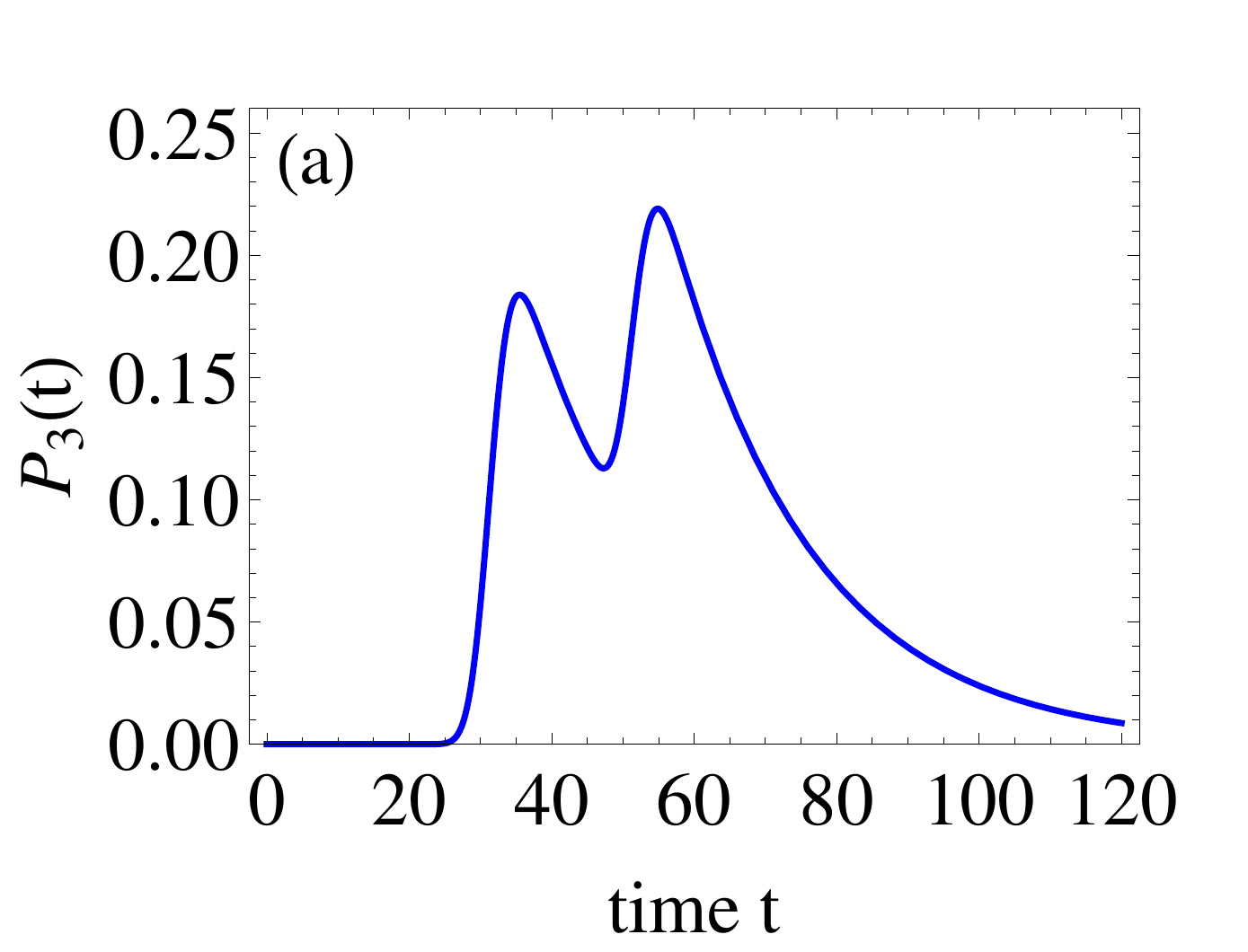}
    \includegraphics[width=0.49\columnwidth]{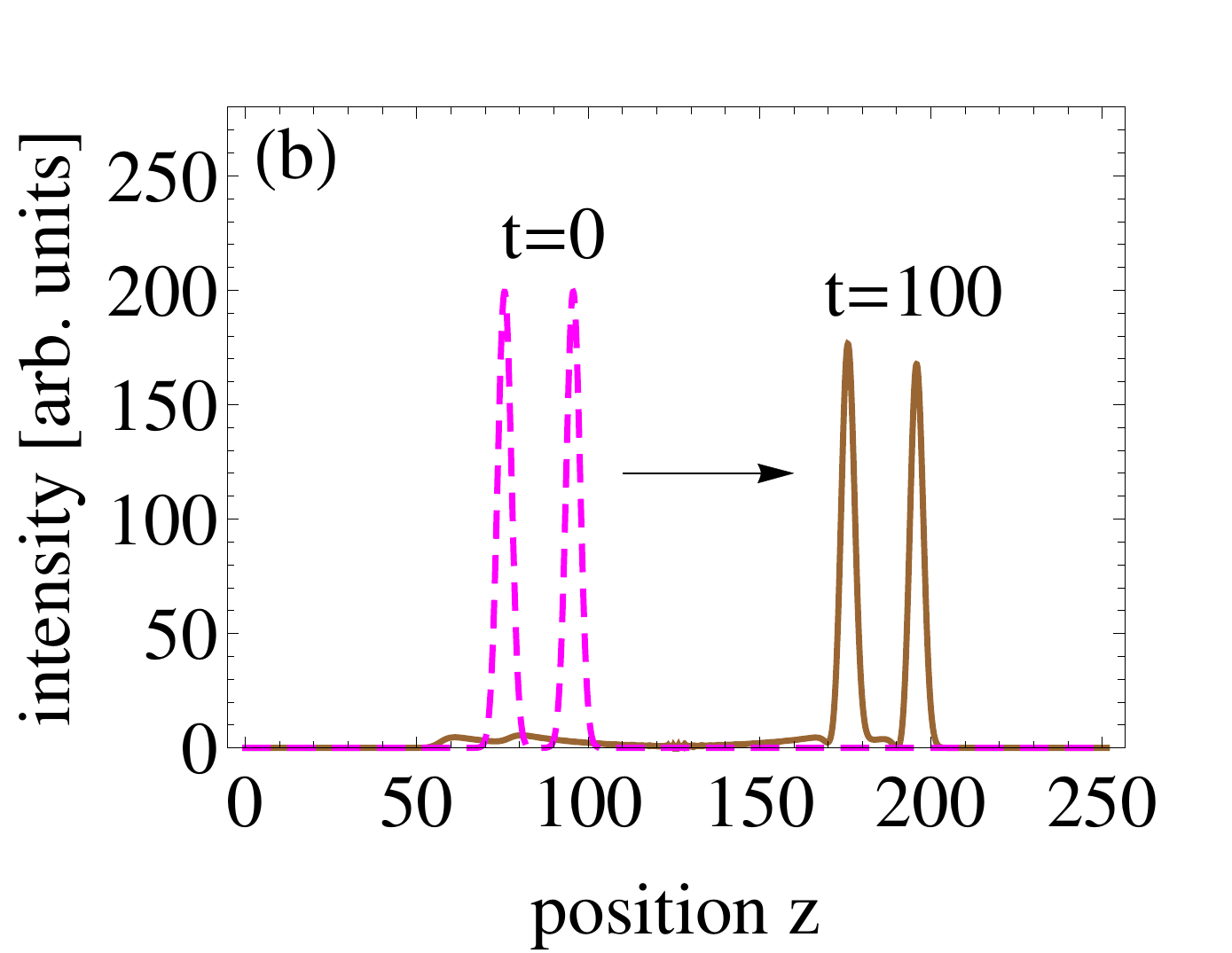}
    \includegraphics[width=0.49\columnwidth]{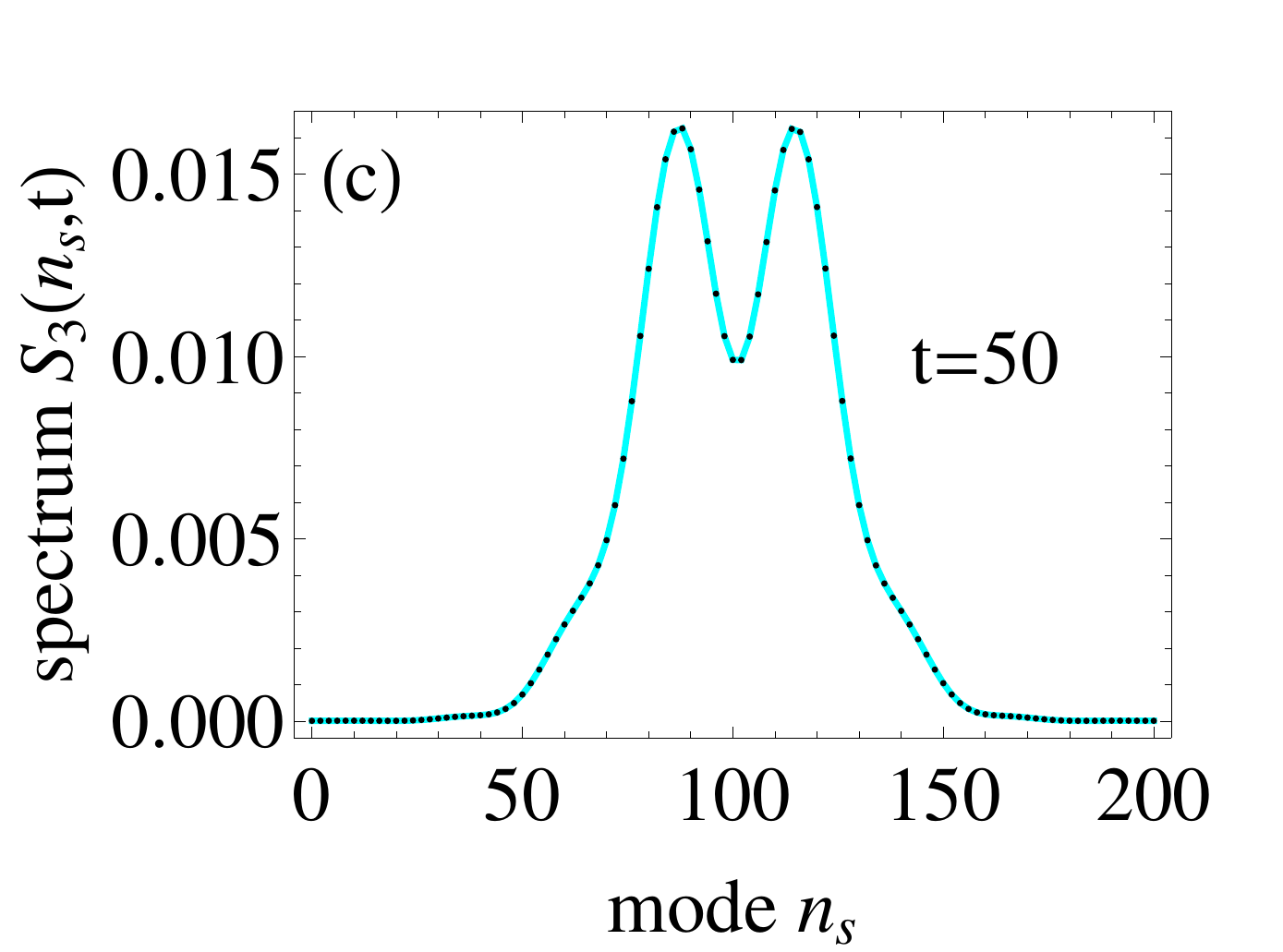}
    \includegraphics[width=0.49\columnwidth]{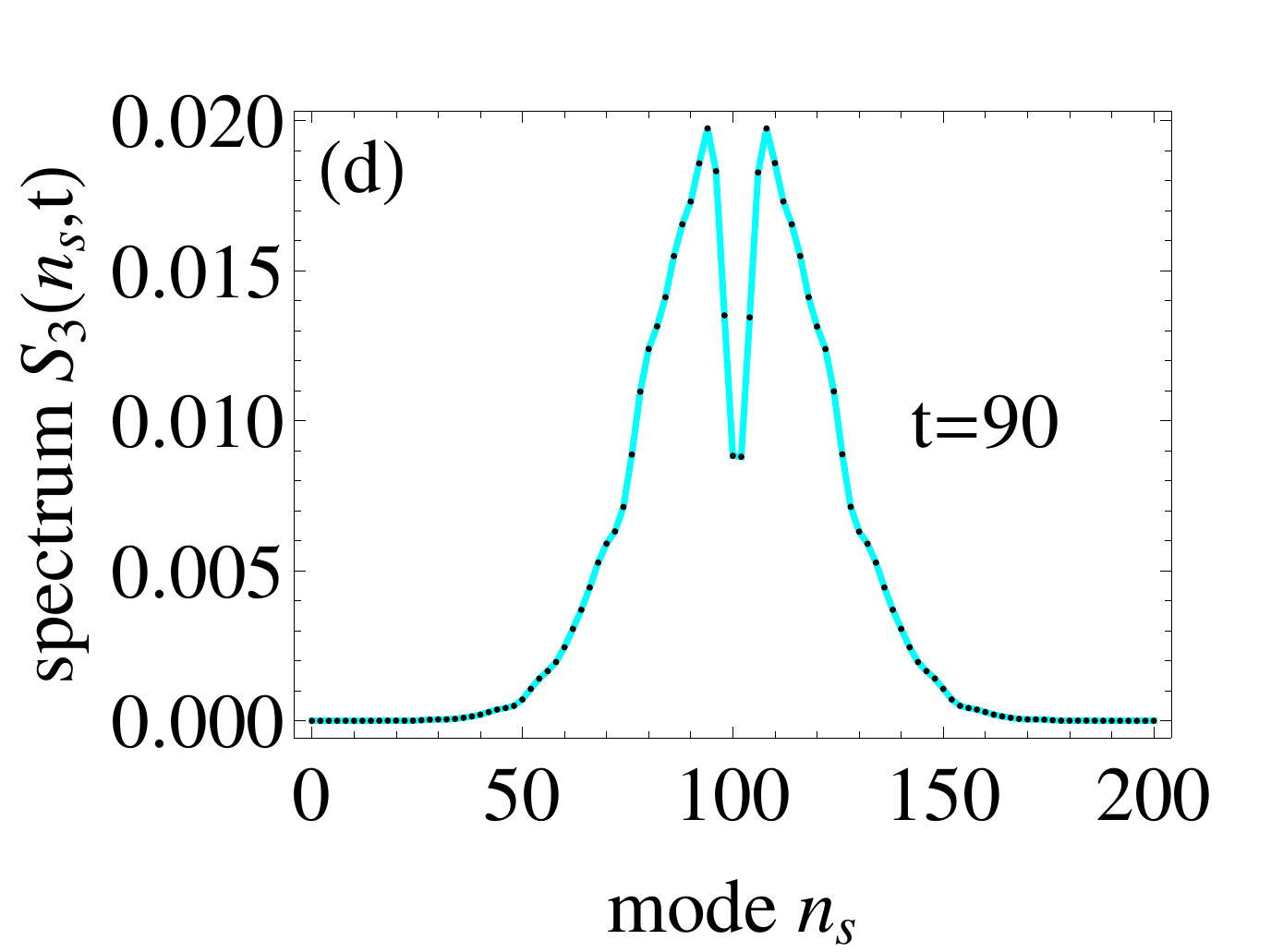}
\caption{(Color online) (a) Time evolution of atomic excitation $P_{3}(t)$ throughout the interaction with two single-photon wave packets. (b) Intensity profile of the two single-photon wave packets before (dashed line, $t=0$) and after (solid line, $t=100$) the interaction. (c) and (d) Cavity mode occupation in momentum space $S_{3}(n_{s},t)$ after interaction with the first pulse ($t=50$) and the second pulse ($t=90$), exhibiting strong absorption of the resonant modes. $n_s$ is defined as in Fig.~\ref{FreeWp} The parameters in dimensionless units ($\hbar=1$, $c=1$) are $z_{1}=95.7$, $z_{2}=75.7$, $L=80\pi$, $\Gamma_{A}=0.05$, $\omega_{0}=\omega_{A}=1000$, $\sigma=0.25$ and $N=200$ modes.}\label{OBESpectrum} 
\end{figure}

Another way of studying stimulated emission is a sequence of two pulses: the first pulse (excitation pulse) leads to the excitation of the atom, whereas the second pulse (stimulating pulse) can cause induced radiation processes. Usually, stimulated emission occurs when a population inversion is present. However, in particular for applications in nuclear quantum optics, we will focus on conditions in which the first pulse is not strong enough to significantly excite the atoms. This will lead to additional interesting phenomena in the interaction with the second pulse.

We choose as initial state with two single photon wave packets at positions $z_{1}$ and $z_{2}$ 
\begin{equation}
\ket{\psi_{3}(0)}= W^{\dagger}(z_{1}) W^{\dagger}(z_{2}) \ket{0} \,,
\end{equation}
which leads to the  initial conditions
\begin{subequations}
\begin{eqnarray}
D_{n}(0)&=& 0  \,, \\ 
E_{n}(0)&=& \frac{1}{\Omega_{N}^{}} \sqrt{2} \left[ G(k_{n}) \right]^{2} e^{-ik_{n}(z_{1}+z_{2})} \,,\\
F_{nm}(0)&=& \frac{1}{\Omega_{N}^{}}  G(k_{n}) G(k_{m}) \left[  e^{-i(k_{m}z_{1}+k_{n}z_{2})} \nonumber \right. \\
 & & \left. + e^{-i(k_{n}z_{1}+k_{m}z_{2})}   \right] \ \text{.}  
\end{eqnarray}
\end{subequations}
The population of the upper atomic level  $P_{3}(t)$ after interacting with the double pulse is shown in Fig.~\ref{OBESpectrum}~(a). Here, the atom is initially in the ground state and the centers of the two wave packets are initially positioned at $z_{1}=95.7$ and $z_{2}=75.7$ and  reach the atom after the times $\Delta t_{1}=30$ and $\Delta t_{2}=50$, respectively. The first single-photon pulse excites the atom to a population of about $0.2$. After the first pulse passed the atom, it decays spontaneously on a slower time scale, because the atomic life time $\tau_{A}=1/\Gamma_{A}$ is a larger than the pulse duration $T_{P}$. The interaction of the second pulse with the already excited atom leads to a higher excitation followed by the spontaneous decay of the atom.

This interpretation is also reflected in the intensity profile shown in Fig.~\ref{OBESpectrum}~(b). After the interaction with the atom at time $t=100$, the height of both pulses is decreased due to the mentioned absorption processes. But the first pulse has decreased more than the second pulse, since the partially excited atom absorbs less radiation than the atoms in the ground state experienced by the first pulse. This can be also seen from the momentum space representation in Fig.~\ref{OBESpectrum}~(c) and (d). After the interaction with the first pulse ($t=50$), the momentum spectrum is modified and we see a minimum in the spectrum which corresponds to strong absorption of the resonant modes by the atom. After the interaction with the second pulse ($t=90$) the absorption profile is centered more narrow around the 
resonant modes and the near-resonant modes are stronger occupied than after the interaction with first pulse. The form of the two spectra consist of superpositions of the Gaussian form of the incoming wave packets and the more narrow Lorentzian form of the spontaneous decay from the partially excited atom. During this two-pulse sequence, no stimulated emission occured because the first pulse was not able to create a population inversion and therefore the second pulse only acts as an excitation pulse and not as a stimulating pulse. However, creating a population inversion is challenging with nuclei in the x-ray regime. Therefore, in the following, we discuss an alternative way of studying stimulated emission.

\subsection{Coherence-enhanced stimulated emission}

In contrast to the usual appearance of stimulated emission as a consequence of an inverted atom interacting with a suitable photon, Drobn\'{y} et al. \cite{BuzekSE} described stimulated emission as a consequence of quantum interference between two wave packets. They were working in the one-excitation subspace, where a suitably double-pulse shaped wave packet first excites the atom, and subsequently de-excites it. Due to the coherent nature of the total process, the impact of the second pulse sensitively depends on the relative phase between the two pulses. For a suitable relative phase, the second pulse can lead to stimulated emission even though the first pulse did not create a population inversion. Their description is not in direct connection to the usual case of stimulated emission, because they only consider one excitation in the system in total. In the following, we will expand the formalism to the two-excitation subspace. 

\begin{figure}[t]
    \includegraphics[width=0.49\columnwidth]{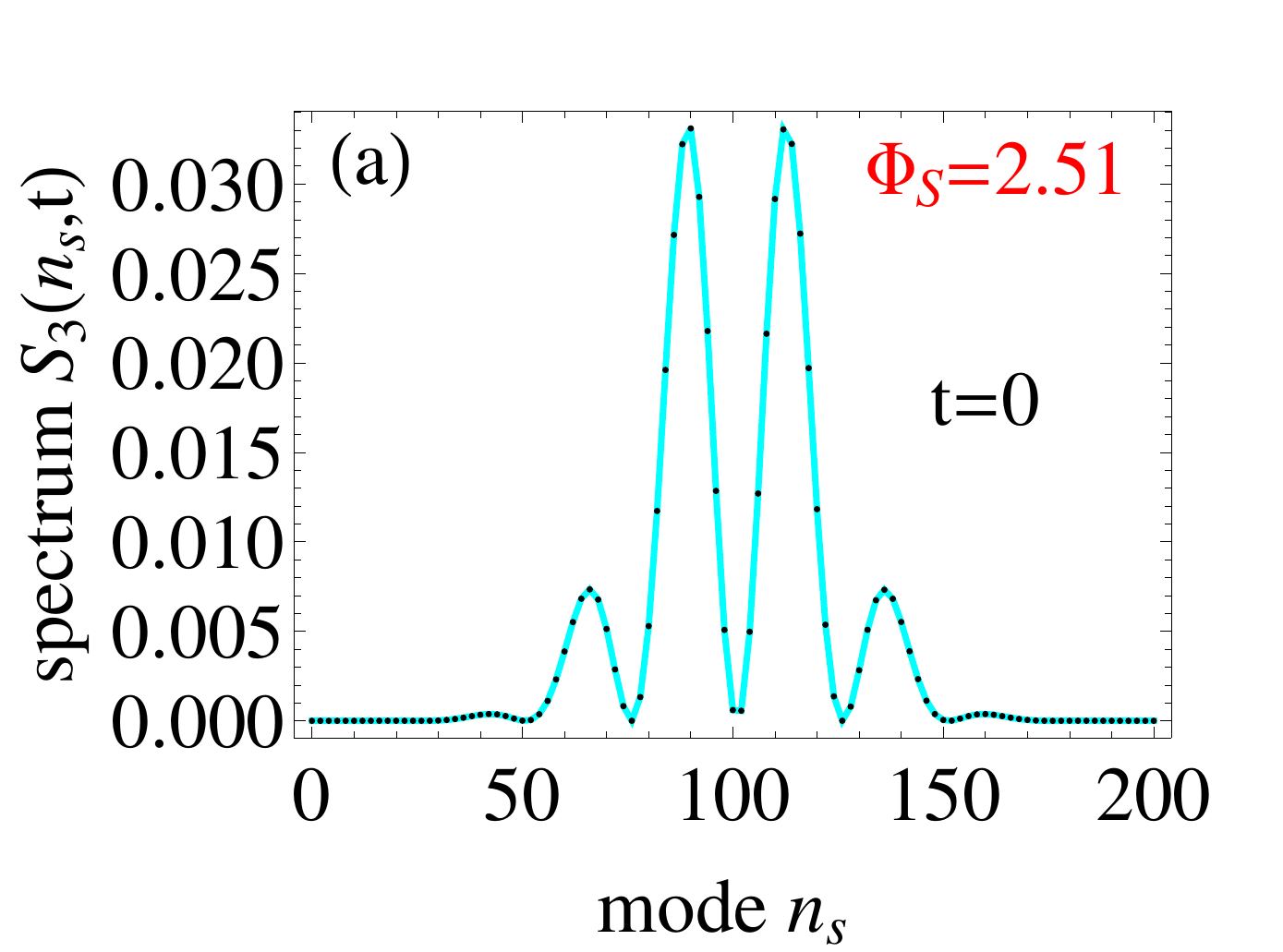}
    \includegraphics[width=0.49\columnwidth]{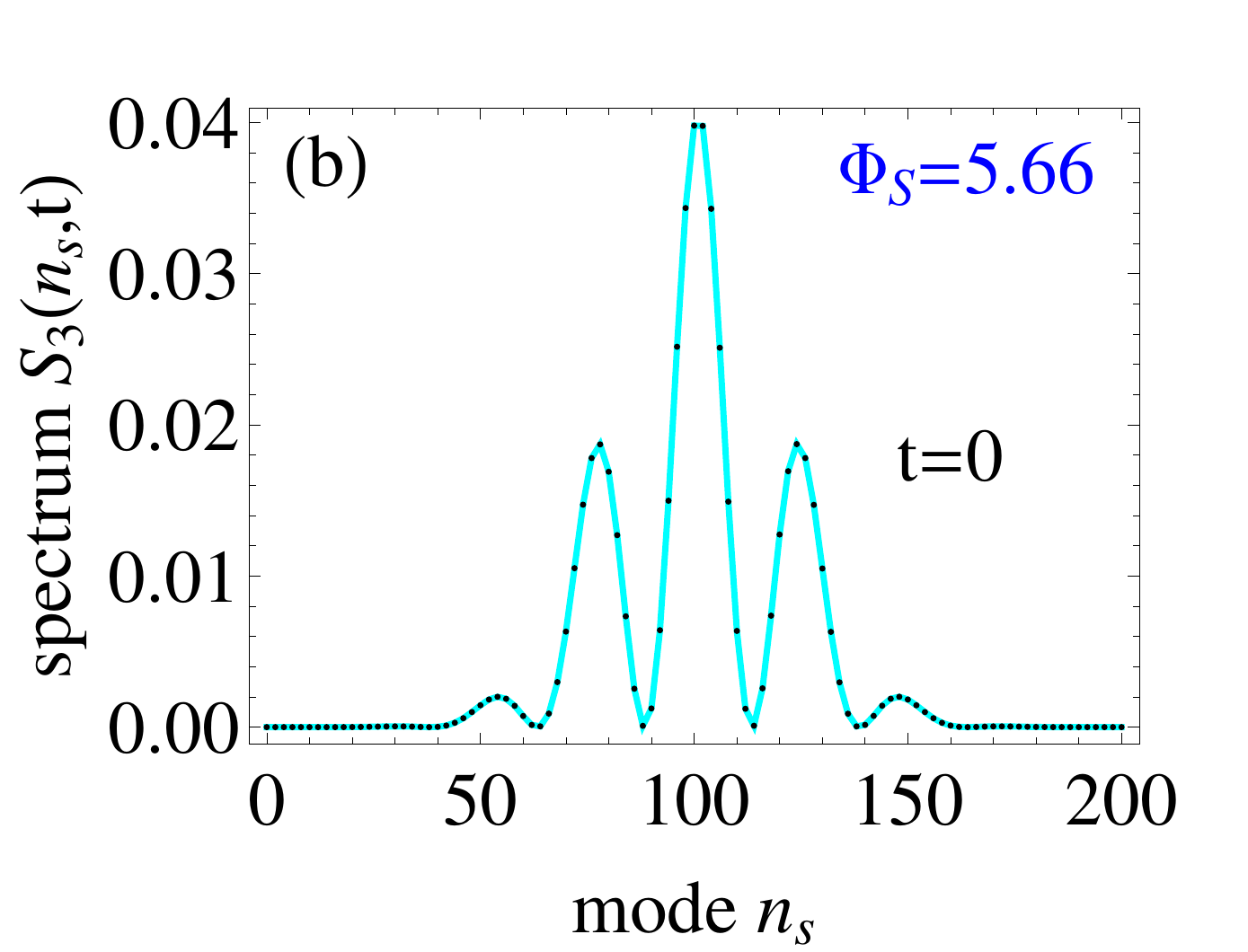}
    \includegraphics[width=0.49\columnwidth]{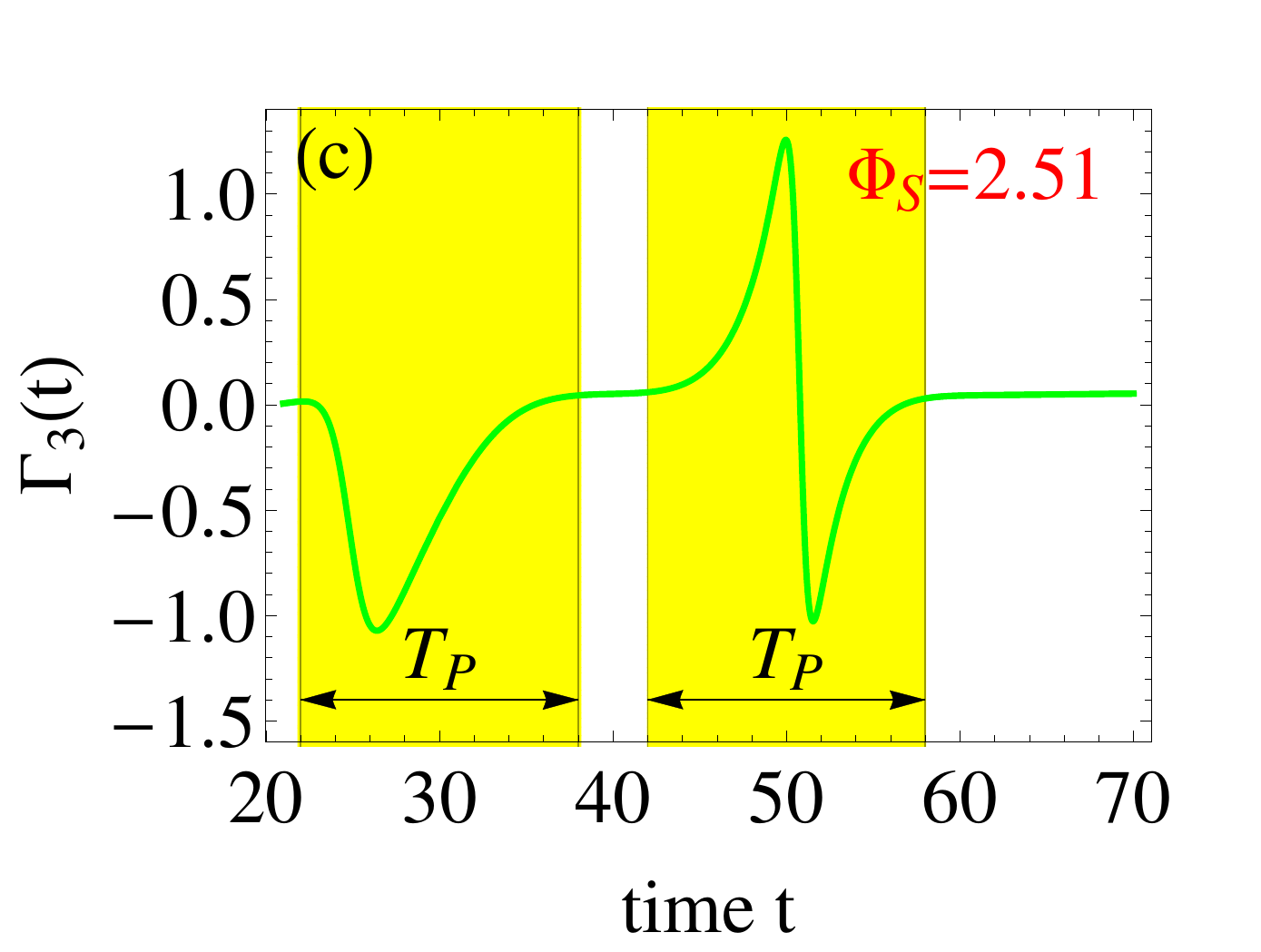}
    \includegraphics[width=0.49\columnwidth]{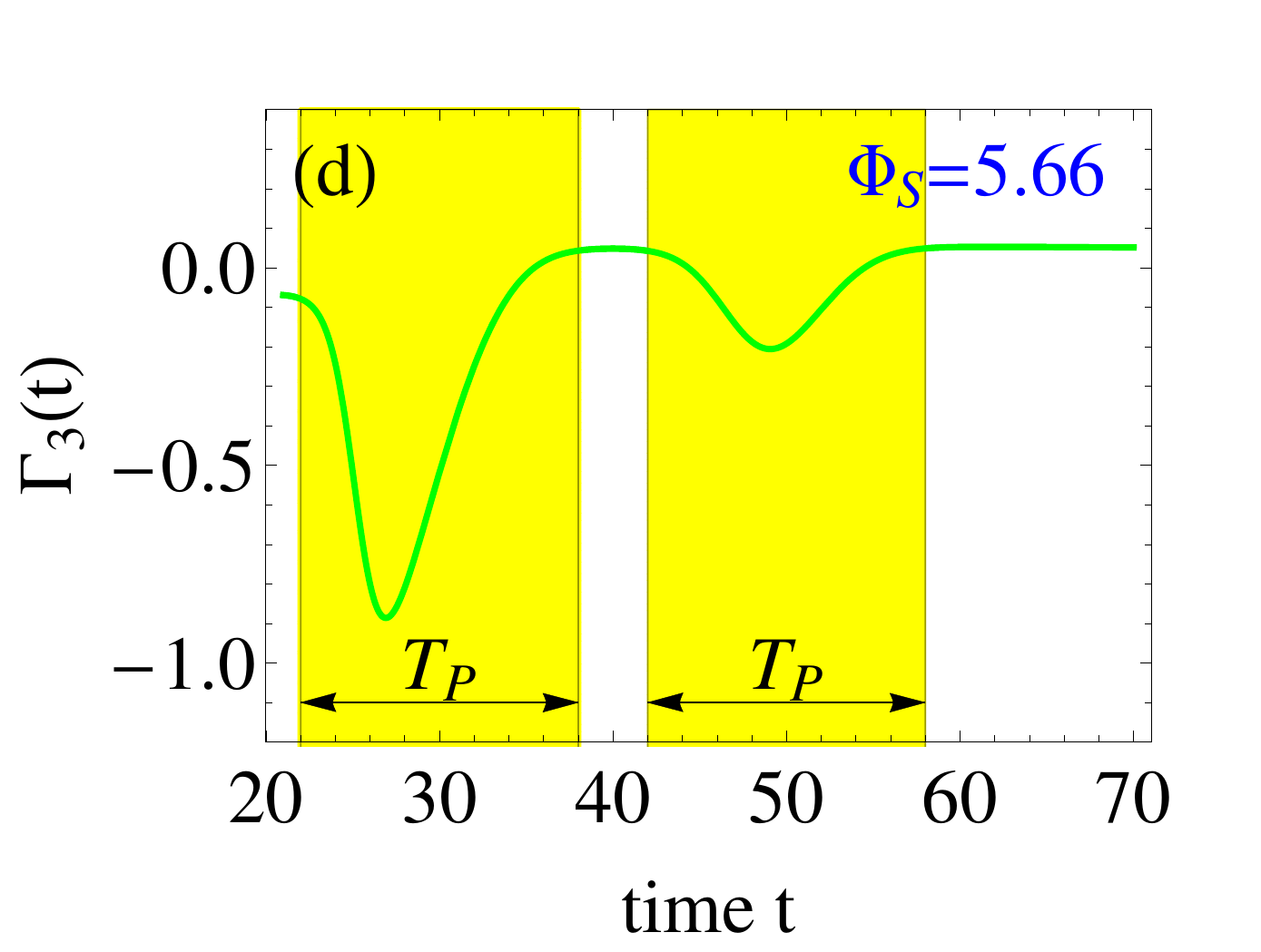}
    \includegraphics[width=0.49\columnwidth]{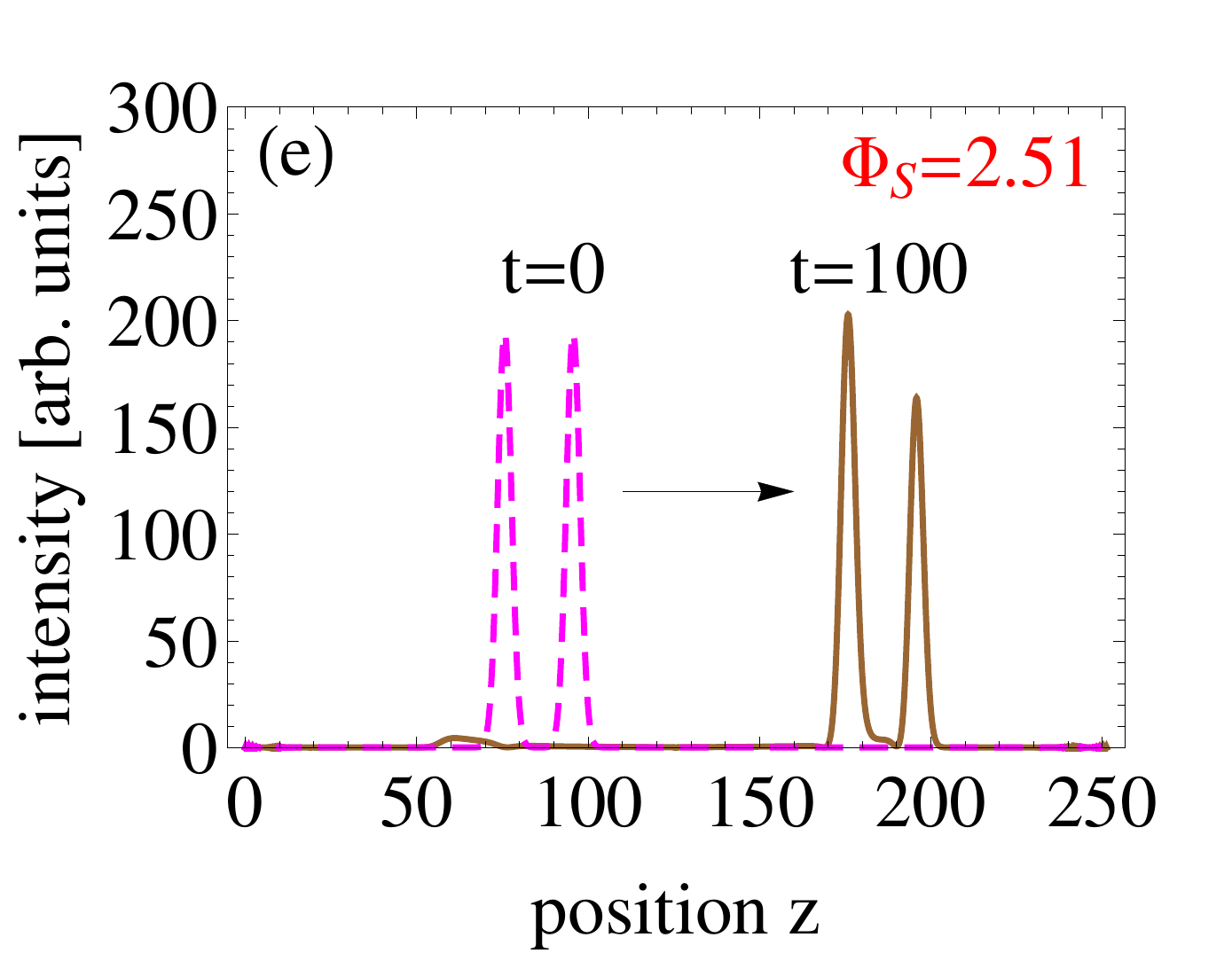}
    \includegraphics[width=0.49\columnwidth]{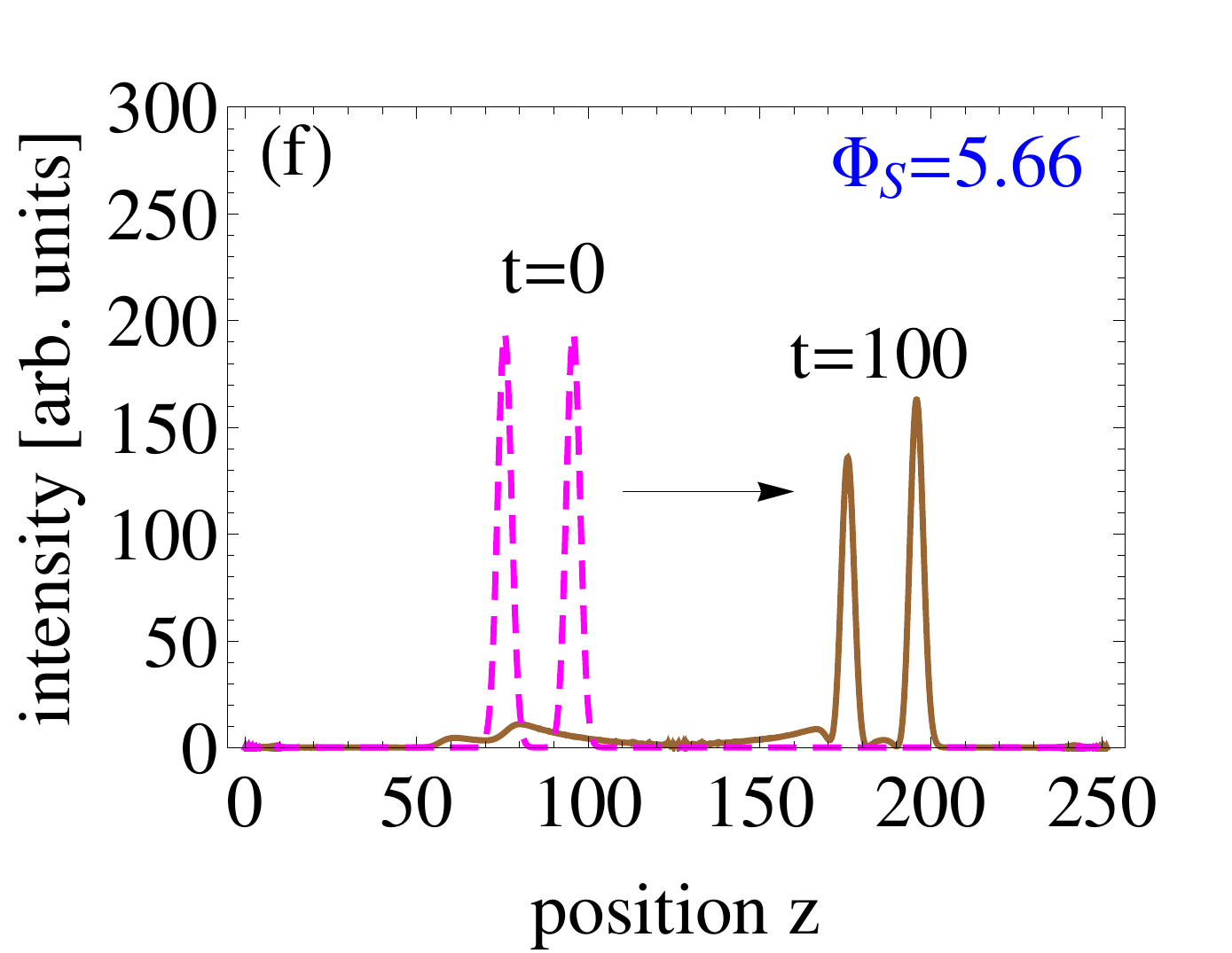}
\caption{(Color online) Initial occupation of the cavity modes in momentum space for relative phases (a) $\Phi_{S}^{min}=2.51$  and (b) $\Phi_{S}^{max}=5.66$. (c,d) show the evolution of the time-dependent decay rate $\Gamma_{3}(t)$ for the two phase values. (e,f) show the corresponding spatial intensity profiles before (t=0, dashed line) and after (t=100, solid line) interaction with the atom. The parameters in dimensionless units ($\hbar=1$, $c=1$) are $z_{1}=95.7$, $z_{2}=75.7$, $L=80\pi$, $\Gamma_{A}=0.05$, $\omega_{0}=\omega_{A}=1000$, $\sigma=0.25$ and $N=200$ modes.}\label{BilderABundSEViaPhase2}
\end{figure}

For creating two photon wave packets with a constant relative phase $\Phi_{S}$ between them, we choose the modified initial state
\begin{equation}
\ket{\psi_{3}(0)}= \frac{1}{N_{\Phi_{S}}}\left[ W^{\dagger}(z_{1}) + e^{i\Phi_{S}} \ W^{\dagger}(z_{2}) \right]^{2} \ket{0} \ \text{,} 
\end{equation}
where we introduced the normalization factor $N_{\Phi_{S}}$. The  initial conditions are
\begin{subequations}
\begin{align}
D_{n}(0)&= 0 \,, \\ 
E_{n}(0)&= \frac{1}{N_{\Phi_{S}}} \frac{1}{\Omega_{N}^{}} \sqrt{2} \left[ G(k_{n}) \right]^{2} \left[ e^{-ik_{n}z_{1}} + e^{-ik_{n}z_{2}+i\Phi_{S}} \right] \,, \\
F_{nm}(0)&= \frac{1}{N_{\Phi_{S}}} \frac{1}{\Omega_{N}^{}}  G(k_{n}) G(k_{m}) \left[ e^{-ik_{n}z_{1}} + e^{-ik_{m}z_{2}+i\Phi_{S}} \right] \nonumber \\
 & \times \left[ e^{-ik_{m}z_{1}} + e^{-ik_{n}z_{2}+i\Phi_{S}} \right] \ \text{.}  
\end{align}
\end{subequations}
\begin{figure}[t]
    \includegraphics[width=0.49\columnwidth]{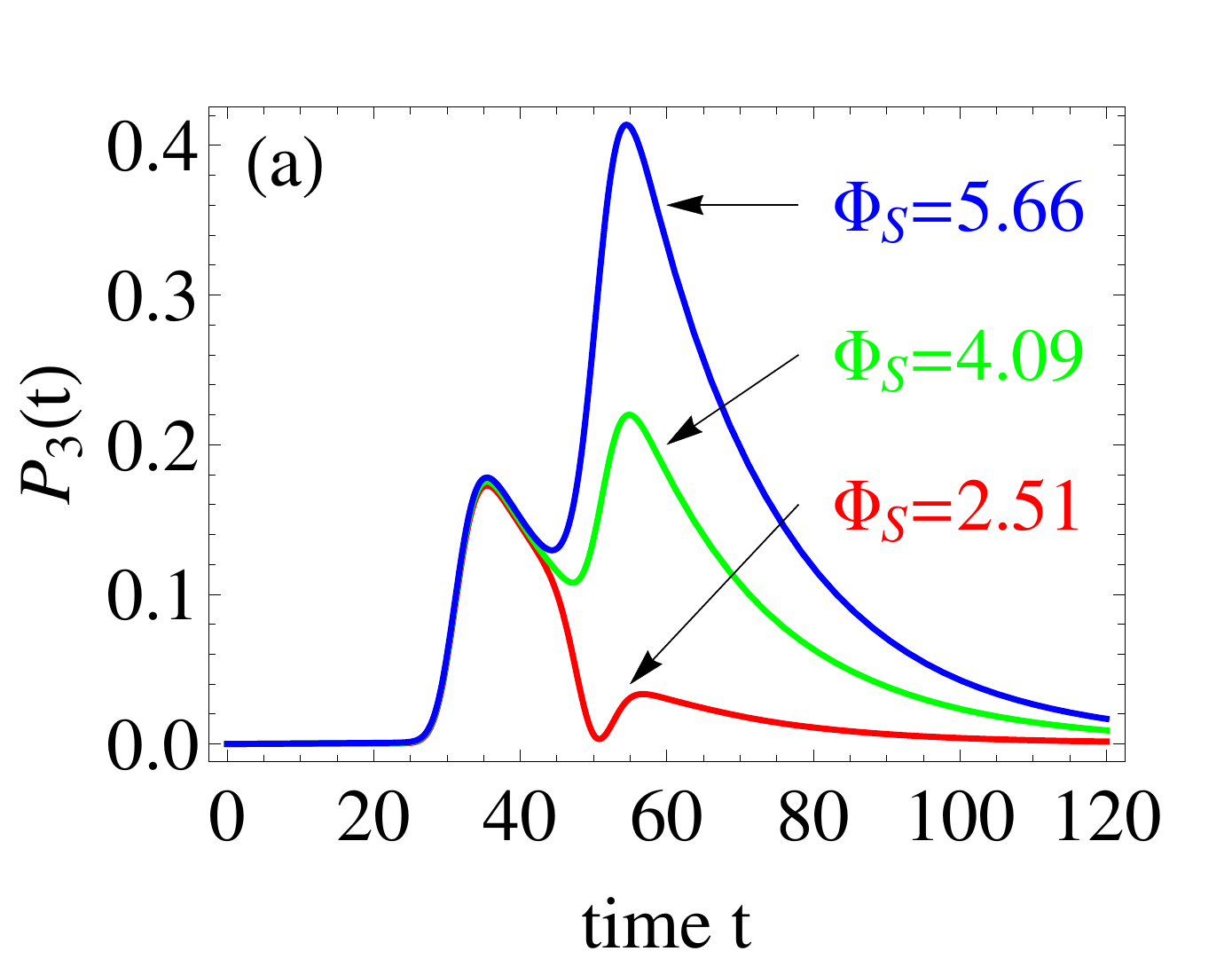}
    \includegraphics[width=0.49\columnwidth]{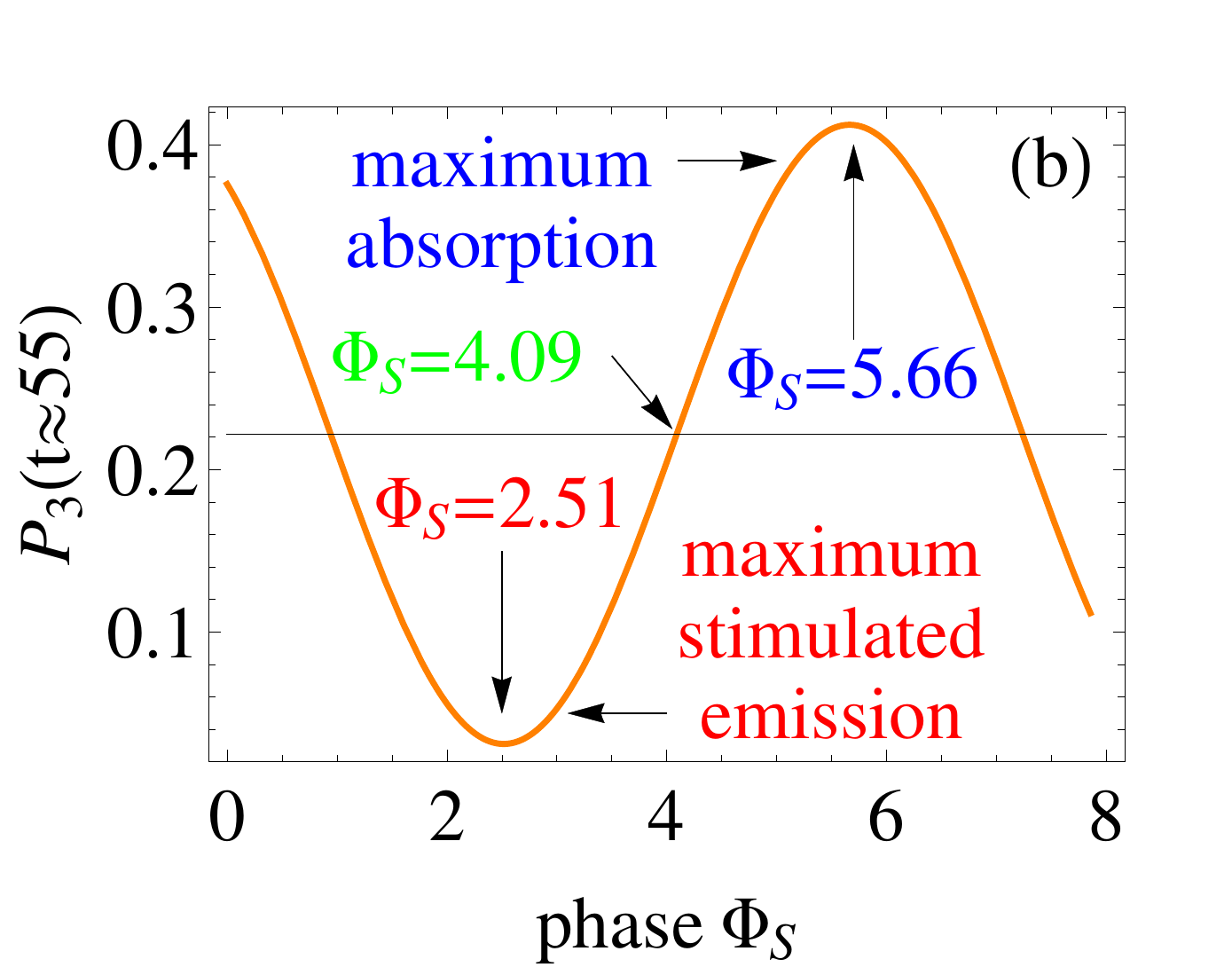}
\caption{(Color online) (a) Atomic excitation $P_{3}(t)$ for excitation with a double pulse with three different relative phases $\Phi_{S}$. (b) Dependence of the maximum of the atomic excitation after the interaction with the second pulse at time $t\approx 55$ as function of the relative phase $\Phi_{S}$. The parameters in dimensionless units ($\hbar=1$, $c=1$) are $z_{1}=95.7$, $z_{2}=75.7$, $L=80\pi$, $\Gamma_{A}=0.05$,  $\omega_{0}=\omega_{A}=1000$, $\sigma=0.25$ and $N=200$ modes.}\label{BilderABundSEViaPhase1}
\end{figure} 
The relative phase $\Phi_{S}$ between the two pulses  sensitively influences the maximum of the atomic excitation after interaction with the second pulse ($t\approx 55$), as it is shown in Fig.~\ref{BilderABundSEViaPhase1}~(a). This dependence is studied in more detail in Fig.~\ref{BilderABundSEViaPhase1}~(b). For the given parameters a maximum of the atomic population occurs at $\Phi_{S}^{max}=5.66$ and a minimum at $\Phi_{S}^{min}=2.51$. 
The effect of the relative phase on the momentum space distribution of the double pulse is illustrated in Fig.~\ref{BilderABundSEViaPhase2}(a,b). Panel (a) shows the case $\Phi_{S}^{min}$, which features a minimum occupation of the resonant mode. In contrast, the case $\Phi_{S}^{max}$ has a maximum occupation of the resonant mode. 

To identify the processes of either absorption or stimulated emission with this two pulse setup, we explore again the time evolution of the decay rate $\Gamma_{3}(t)$ like in Sec.~\ref{ChapIntSE}. As expected, the interaction with the first pulse shows the same absorption profile for both extrema, see Fig.~\ref{BilderABundSEViaPhase2} (c) and (d). 

For $\Phi_{S}^{min}$ in Fig.~\ref{BilderABundSEViaPhase2}(c), the second pulse leads to a dispersive shape of the time-dependent decay rate. Initially, the rate increases to positive values, which can be interpreted as stimulated emission. Subsequently, the rate turns negative, indicating a reabsorption of the atom. It is interesting to note that the enhanced decay rate has a maximum which exceeds the value given by Fermi's golden role by a factor of 25. This is much higher than the maximal enhancement during stimulated emission via population inversion~\cite{Shanhui,Valente2}. In contrast, for $\Phi_{S}^{max}$ in Fig.~\ref{BilderABundSEViaPhase2}(d),  also the second pulse leads to absorption indicated by a negative rate throughout the entire pulse.

Additional signatures of absorption and stimulated emission can be found in position space, by regarding the intensity profiles at two different times: before and after interaction with the atom. After the interaction at time $t=100$, both plots show the same decrease of the front pulse, due to absorption of radiation by the atom. But for the phase $\Phi_{S}^{min}$, we see a strong increase of the second stimulating pulse in Fig.~\ref{BilderABundSEViaPhase2}~(e), which is because of stimulated emission of radiation into the direction of the stimulating pulse during the pulse duration. On the other hand for $\Phi_{S}^{max}$ in Fig.~\ref{BilderABundSEViaPhase2}~(f), the second excitation pulse is even smaller than the front pulse, which reflects the strong absorption due to 
constructive quantum interference.

\section{Semiclassical description of stimulated emission}

For the study of nuclear parameters, the excitation pulses are usually very short in comparison to the atom's lifetime. We will consider the iron isotope ${}^{57}\text{Fe}$ as nuclear target, 
because it provides a recoilless resonant M\"ossbauer transition \cite{Moessbauer} and has been facilitated in most experiments on nuclear quantum optics up to now. The transition energy is $14.4$ keV and the nuclei have a lifetime of $141$ ns. The usual duration of the excitation pulses are in the order of ps- to fs-range, depending on the light source. This corresponds to a difference of at least three orders of magnitude between the relevant time scales which leads to high demands on the numerical implementation, since a large number of modes has to be taken into account in order to ensure convergence. Another issue is that our model is restricted to 1D because of the computational effort required for higher-dimensional simulations, while the 1D setup is sufficient for revealing temporal aspects of stimulated emission. In the following, we will therefore switch to a semiclassical description with a 3D Wigner-Weisskopf decay rate, where the electromagnetic field is traced out and the numerical modeling becomes substantially easier to manage. As a first step, we will discuss the relation between the quantum and the semiclassical descriptions.

\subsection{Accordance with unitary 1D-cavity model}

The following description is reminiscent of an earlier work on the semiclassical description of a single-photon wave packet interacting with an excited atom \cite{Elyutin2012}, but extends it to a two-pulse sequence. The electric field is chosen as a linearly polarized pulse with a Gaussian shape, propagating in the positive $z$ direction 
\begin{eqnarray}
\textbf{E}(z,t)=\textbf{e}_{x} \mathcal{E}_{0} &\Bigl\{& \exp{\left[ -(z-\Lambda_{1}c-tc)^2\sigma^{2} \right]} \nonumber \\ & & \times \cos\left[k_{0}(z-\Lambda_{1}c)-\omega_{0}t\right] \nonumber \\ &+& \exp{\left[ -(z-\Lambda_{2}c-tc)^2\sigma^{2} \right]} \nonumber \\
& & \times \cos\left[k_{0}(z-\Lambda_{2}c)-\omega_{0}t\right] \Bigr\} \ \text{,}
\end{eqnarray}
where $\mathcal{E}_{0}$ is the amplitude of the pulse, $\textbf{e}_{x}$ is the polarization vector, $\sigma$ is the dispersion of the wave packet, $k_{0}=\omega_{0}/c$ is the wave number, $c$ is the speed of light. Finally, $\Lambda_{1}$ as well as $\Lambda_{2}$ are the temporal displacements of the two wave packet centers from the atom position. The amplitude $\mathcal{E}_{0}$ can be fixed via considering the total energy of the field \cite{Jackson}
\begin{equation}\label{Energy_EM}
\frac{1}{2} \int_{V}  \left[ \varepsilon_{0}\textbf{E}^{2}(\textbf{r},t)+ \frac{1}{\mu_{0}} \textbf{B}^{2}(\textbf{r},t)\right] \mathrm{d}V = n_{res} \cdot 2 \hbar \omega_{0}
\end{equation}
to be equal to the number of resonant photons $n_{res}$ in each pulse. We have $\mathrm{d}V=\mathrm{d}z\cdot \mathrm{d}A_{beam}$, where we assumed that the field is restricted to an area in the transverse direction which corresponds to the beam size $A_{Beam}$ of a light source. The impact of the two pulses on a two-level atom can be described by using the master equation/optical Bloch equations in the interaction picture~\cite{Scully,Ficek,Kiffner_Vacuum_Processes}
\begin{equation}
\frac{\partial}{\partial t} \rho(t) = - \frac{i}{\hbar} \left[{\hat H}^{}_{int}(t),\rho(t) \right] + \mathcal{L}_{sd}\left[\rho(t)\right] \ \text{,}
\end{equation}
where $\mathcal{L}_{sd}$ describes the coupling of the atom to its environment and is given by \cite{LindbladSuper}
\begin{equation}
\mathcal{L}_{sd}\left[\rho(t)\right]=
\begin{pmatrix}
\Gamma_{A}\rho_{22}(t) & -\frac{1}{2}\Gamma_{A}\rho_{12}(t)  \\
-\frac{1}{2}\Gamma_{A}\rho_{21}(t)  & -\Gamma_{A}\rho_{22}(t)
\end{pmatrix} \ \text{.}
\end{equation}
The atom-field Hamiltonian in the interaction picture is
\begin{equation}
{\hat H}^{}_{int}(t)=
\begin{pmatrix}
0 & \frac{\hbar}{2} \Omega_{}(t)  \\
\frac{\hbar}{2} \Omega_{}^{*}(t) & - \hbar \Delta_{} 
\end{pmatrix} \ \text{,}
\end{equation}
with the the time-dependent Rabi frequency 
\begin{equation}\label{Rabi1}
\Omega_{}(t)=\frac{d_{eg}E_{0}(t)}{\hbar} \ \text{,}
\end{equation}
where $d_{eg}$ ist the 1D Wigner-Weisskopf rate from Eq.~(\ref{WWRate1D}). We assume resonant interaction with the light pulses and therefore  $\Delta=\omega_{A}-\omega_{0}=0$ in the following. The envelope is
\begin{eqnarray}
E_{0}(t)=\mathcal{E}_{0} &\Bigl\{& \exp{\left[ -(t-\Lambda_{1})^2(\sigma c)^{2} \right]} \nonumber \\
&+& \exp{\left[ -(t-\Lambda_{2})^2(\sigma c)^{2} +i\Phi_{M} \right]} \Bigr\} \ \text{,}
\end{eqnarray}
\begin{figure}[t]
    \includegraphics[width=0.7\columnwidth]{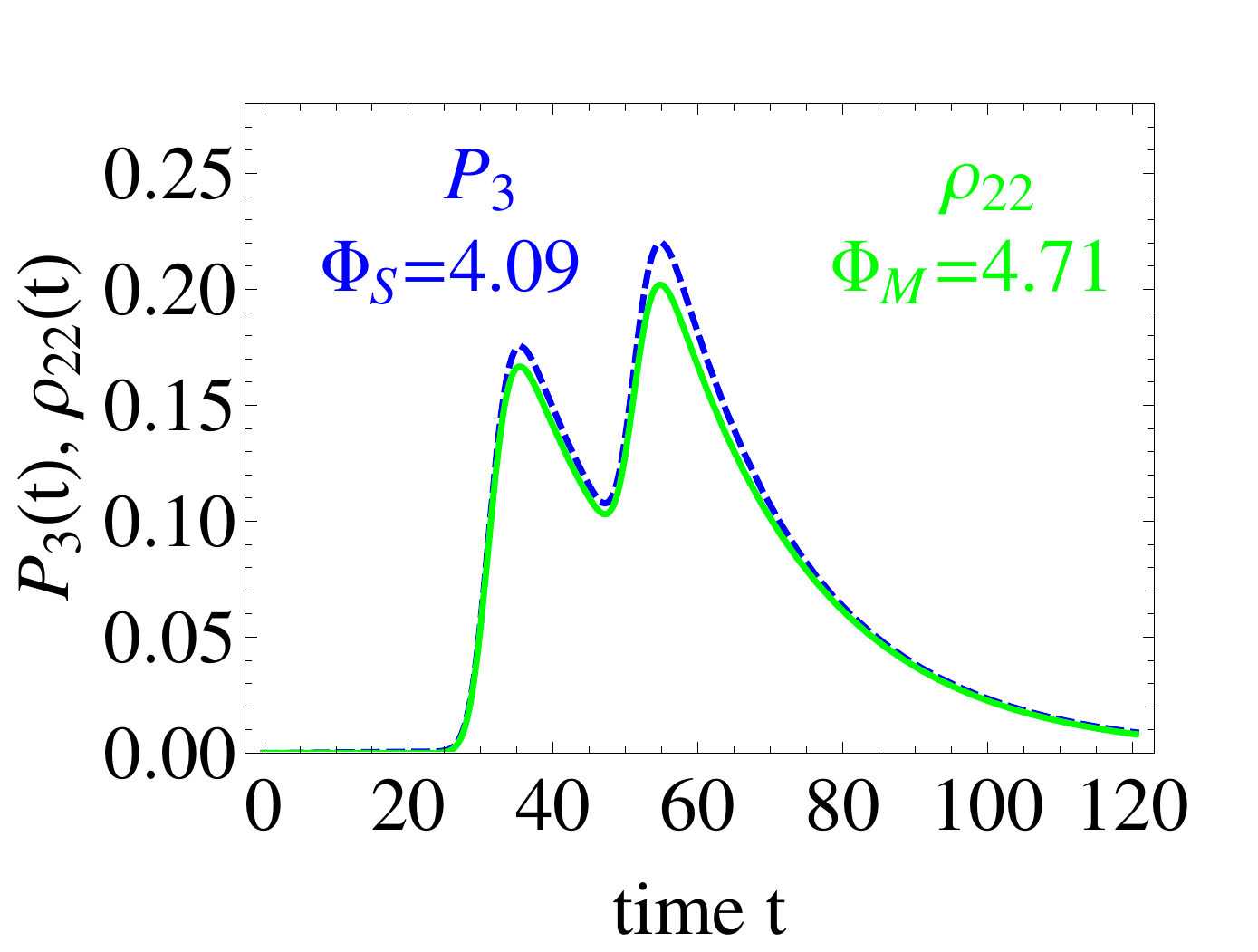}
\caption{(Color online) Time evolution of the atomic excitations $P_{3}(t)$ with phase $\Phi_{S}=4.09$ and $\rho_{22}(t)$ with phase $\Phi_{M}=4.71$. The phase difference $\Phi_{M}-\Phi_{S}$ corresponds to $\Delta\Phi=(\Lambda_{2}-\Lambda_{1})k_{0}c$ up to multiples of $2\pi$ and is due to the different definition of $\Phi_{S}$ and $\Phi_{M}$. The parameters in dimensionless units ($\hbar=1$, $c=1$) are $z_{1}=95.7$, $z_{2}=75.7$, $\Lambda_{1}=30$, $\Lambda_{2}=50$, $L=80\pi$, $\Gamma_{A}=0.05$, $\omega_{0}=\omega_{A}=1000$, $\sigma=0.25$, $A_{Beam}=1$, $n_{res}=1$ and $N=200$ modes.}\label{BeziehungPhasen}
\end{figure} 
where we inserted a relative phase $\Phi_{M}$ between the two pulses. The time delay between the pulses is $\tau_{d}=\Lambda_{2}-\Lambda_{1}$.
In Fig.~\ref{BeziehungPhasen}, we compare the results of the Schr\"odinger equation calculation [$P_{3}(t)$] with those of the master equation [$\rho_{22}(t)$]. Qualitatively, the two results agree well,  up to a relative phase shift between the phases $\Phi_{S}$ and $\Phi_{M}=\Delta\Phi+\Phi_{S}$ in the two models given by $\Delta\Phi=(\Lambda_{2}-\Lambda_{1})k_{0}c$ which is due to the fact that phase $\Phi_{S}$ is introduced in momentum space, whereas $\Phi_{M}$ is introduced in the time domain. 
Quantitatively, the maximum atomic excitation predicted by the two models does not fully coincide. One reason is that the Schr\"odinger equation relies on a discrete set of photon modes, whereas the master equation is obtained by integrating over a continuum of modes.  Consequently, the accordance of  the two models becomes better with pulse durations much shorter than the atomic lifetime: $T_{P}\ll\tau_{A}$.

\subsection{Semiclassical analysis of stimulated emission}

For analyzing coherence-enhanced stimulated emission in a two pulse setup, we rewrite the master equation into a vector-matrix representation
\begin{equation}
\frac{\mathrm{d}}{\mathrm{d}t}\boldsymbol{\rho}(t)= \mathcal{M}(t) \boldsymbol{\rho}(t)
\end{equation}
with the density-matrix vector
\begin{equation}
\boldsymbol{\rho}(t)= \Bigl(\rho_{11}(t),\rho_{12}(t),\rho_{21}(t),\rho_{22}(t)\Bigr)^{T}
\end{equation}
and the interaction matrix 
\begin{equation}
\mathcal{M}(t) = \begin{pmatrix} 0 & \frac{i\Omega(t)}{2} & -\frac{i\Omega^{*}(t)}{2} & \Gamma_{A} \\
			      \frac{i\Omega^{*}(t)}{2} & -\frac{\Gamma_{A}}{2} & 0 & -\frac{i\Omega^{*}(t)}{2} \\
			      -\frac{i\Omega(t)}{2} &0 &  -\frac{\Gamma_{A}}{2} & \frac{i\Omega(t)}{2} \\
			      0 & -\frac{i\Omega(t)}{2} & \frac{i\Omega^{*}(t)}{2} & -\Gamma_{A} \\ 
    \end{pmatrix} \ \text{.}
\end{equation}
This representation is more useful for approximative expressions and its formal solution is given by
\begin{eqnarray}\label{Taylor}
\boldsymbol{\rho}(t) &=& e^{\int_{0}^{t}\mathcal{M}(t')\mathrm{d}t'} \boldsymbol{\rho}(0) = \left[ \mathds{1} + \int_{0}^{t}\mathcal{M}(t')\mathrm{d}t'\right.  \nonumber \\
& &  \left. + \int_{0}^{t}\mathcal{M}(t')\mathrm{d}t' \int_{0}^{t'}\mathcal{M}(t'')\mathrm{d}t''  + \ldots \right] \boldsymbol{\rho}(0)  \ \text{.} \nonumber 
\end{eqnarray}
We expanded this solution to the second order, which is a valid approximation in the low excitation regime which will be relevant for nuclear parameters. 

Because of our analysis in the first part we know that the stimulated photons will be emitted within the interaction time with the second stimulating pulse. Therefore, we calculate the effect of the first pulse and the subsequent spontaneous emission using the full master equation, and use the Taylor expansion in Eq.~(\ref{Taylor}) only to describe the interaction with the second pulse. To isolate the several effects (spontaneous decay, stimulated emission and absorption) the second pulse will have on the atomic dynamic, we  analyze the density matrix elements $\rho_{ij}(t)$ sequentially at different times. This requires a new envelope only containing one pulse 
\begin{equation}\label{NewEnvelope}
E_{0}(t)=\mathcal{E}_{0}  \exp{\left[ -(t-\Lambda_{1})^2(\sigma c)^{2} \right]} \ \text{.}
\end{equation}
Starting from $t=0$, after the interaction with the first pulse, the atom spontaneously decays before the second pulse reaches the atom at time $t_{int}$. As example, we choose the time between the pulses such that the atom looses $10\%$ of its excitation due to spontaneous emission. To identify the different processes during the interaction with the second pulse, we then use the expansion in Eq.~(\ref{Taylor}) but with initial vector $\boldsymbol{\rho}(t=t_{int})$ obtained from the evolution after the first pulse, and with envelope Eq.~(\ref{NewEnvelope}) multiplied with the relative phase factor $e^{i\Phi_{M}}$. The vector-matrix multiplication leads to the following expressions for the population of the ground and excited state after interacting with the second pulse: 
\begin{subequations}
\begin{align}
\rho_{11}(t=t_{int}+T_{P}) \approx & \ \rho_{11}(t=t_{int}) + \mathcal{R}_{sd} + \mathcal{R}_{se} \nonumber \\ & - \mathcal{R}_{ab} + \mathcal{R}_{\Phi_{1}} - \mathcal{R}_{\Phi_{2}}  \,,\\
\rho_{22}(t=t_{int}+T_{P}) \approx  & \ \rho_{22}(t=t_{int}) - \mathcal{R}_{sd} - \mathcal{R}_{se} \nonumber \\ & + \mathcal{R}_{ab} - \mathcal{R}_{\Phi_{1}} + \mathcal{R}_{\Phi_{2}} \ \text{.}
\end{align}
\end{subequations}
In detail, the term which describes the spontaneous decay during the interaction is
\begin{equation}
\mathcal{R}_{sd}= \left[T_{P}\Gamma_{A} - \frac{\Gamma_{A}^{2}T_{P}^{2}}{2} \right] \rho_{22}(t=t_{int}) \ \text{,}
\end{equation}
because it decreases the excited and increases the ground state by the same amount and it does not couple to the light field, which would be indicated by a multiplication with the Rabi frequency $\Omega(t)$. Stimulated emission is described by the expression
\begin{eqnarray}
\mathcal{R}_{se}&=& \frac{1}{2} \left[\int_{t_{int}}^{t_{int}+T_{P}}\mathrm{d}t'\operatorname{Re}\left[\Omega(t')\right] \int_{t_{int}}^{t'}\mathrm{d}t''\operatorname{Re}\left[\Omega(t'')\right]\right] \nonumber \\ & & \rho_{22}(t=t_{int}) \ \text{,}
\end{eqnarray}
because this term decreases the excited and increases the ground state by the same amount, which is a consequence of the coupling to the light field as indicated by the presence of the Rabi frequency $\Omega(t)$. The term responsible for absorption during the interaction with the pulse is
\begin{eqnarray}
\mathcal{R}_{ab}&=& \frac{1}{2} \left[\int_{t_{int}}^{t_{int}+T_{P}}\mathrm{d}t'\operatorname{Re}\left[\Omega(t')\right] \int_{t_{int}}^{t'}\mathrm{d}t''\operatorname{Re}\left[\Omega(t'')\right]\right] \nonumber \\ & & \rho_{11}(t=t_{int}) \ \text{,}
\end{eqnarray}
because it has the reversed effect of stimulated emission. 

The two most important terms responsible for the coherent enhancement of stimulated emission are those which posses a phase-dependence
\begin{subequations}
\begin{align}
\mathcal{R}_{\Phi_{1}}=&\cos(\Phi_{M})\left[\int_{t_{int}}^{t_{int}+T_{P}} \mathrm{d}t'\int_{t_{int}}^{t'}\mathrm{d}t''\operatorname{Re}\left[\Omega(t'')\right] \right. + \nonumber \\ 
&  \left.\frac{1}{2}\int_{t_{int}}^{t_{int}+T_{P}} \mathrm{d}t'\int_{t_{int}}^{t'}\mathrm{d}t''\operatorname{Re}\left[\Omega(t')\right] \right] \nonumber \\
&  \times \operatorname{Im}\left[\rho_{12}(t=t_{int})\right] \cdot \Gamma_{A}  \nonumber \\
\mathcal{R}_{\Phi_{2}}=&\cos(\Phi_{M}) \left[\int_{t_{int}}^{t_{int}+T_{P}}\operatorname{Re}\left[\Omega(t')\right]\mathrm{d}t'\right] \nonumber \\ 
&  \times \operatorname{Im}\left[\rho_{12}(t=t_{int})\right] \text{.}
\end{align}
\end{subequations}
According to the choice of the phase $\Phi_{M}$, the cosine will change its sign and therefore these terms are either responsible for stimulated emission or responsible for absorption. For example, in the interval $\Phi_{M}\in[\tfrac{\pi}{2}, \tfrac{3\pi}{2}]$ the term $\mathcal{R}_{\Phi_{1}}$ describes absorption and $\mathcal{R}_{\Phi_{2}}$ describes stimulated emission whereas for the intervals $\Phi_{M}\in[0, \tfrac{\pi}{2}]$ and $\Phi_{M}\in[\tfrac{3\pi}{2},\tfrac{5\pi}{2}]$ it is the other way around. It should be noted that the two phase-dependent terms differ in magnitude, because $\mathcal{R}_{\Phi_{1}}$ is proportional to $T_{P}\Gamma_{A}$. This means we can evoke a discrimination between the order of magnitude of these two terms, when we for example use nuclear parameters. There the pulse duration is much shorter than the life time of a nucleus and therefore $|\mathcal{R}_{\Phi_{2}}|$ would exceed $|\mathcal{R}_{\Phi_{1}}|$ significantly. This allows us to use the 
phase interval $\Phi_{M}\in[\tfrac{\pi}{2}, \tfrac{3\pi}{2}]$ as preferred region for observing stimulated emission, because it clearly exceeds absorption.
Because the stimulated photons will be emitted during interaction with a photon pulse, as already discussed, the event rates for the number of stimulated photons $N_{se}$ can be approximated for the respective intervals of the phase
\begin{align}\label{Nse}
N_{se} =
  \begin{cases}
   \mathcal{R}_{se} + \left| \mathcal{R}_{\Phi_{1}} \right| & \text{if } \Phi_{M}\in[0, \tfrac{\pi}{2}] \ \cup [\tfrac{3\pi}{2} ,\tfrac{5\pi}{2}]\\
   \mathcal{R}_{se} + \left| \mathcal{R}_{\Phi_{2}} \right| & \text{if } \Phi_{M}\in[\tfrac{\pi}{2}, \tfrac{3\pi}{2}]
  \end{cases}\,.
\end{align}

\section{Estimation of the event rate of stimulated emission in nuclei with x-ray pulses}

\subsection{Broadband excitation of nuclear targets}

For the excitation of nuclear media, short x-ray pulses in the time domain are used which possess a broadband spectrum in the frequency domain. Therefore only a  small fraction of the photons in the pulse will hit the nuclear resonance width and the other non-resonant photons will pass the nuclear target without exciting it. To take this into account the calculation of the Rabi frequency has to be modified~\cite{PRLEvers06}. The electromagnetic field amplitude of a single x-ray pulse in the time domain can be described by
\begin{equation}
E_{0}(t)=\mathcal{E}_{0} e^{-\frac{t^2}{2\sigma^{2}_{t}}}e^{-i\omega_{0}t} \ \text{,}
\end{equation}
where $\sigma_{t}$ is the width of the wave packet in the time domain. For calculating $\mathcal{E}_{0}$ it is easier to switch into the frequency representation via calculating the Fourier transform 
\begin{equation}
\hat{E}_{0}(\omega)=\frac{1}{\sqrt{2\pi}} \int_{-\infty}^{\infty}E_{0}(t)e^{-i\omega t} \mathrm{d}t 
=\frac{\mathcal{E}_{0}}{\sigma_{\omega}}e^{-\frac{(\omega-\omega_{0})^{2}}{2\sigma_{\omega}^{2}}} \ \text{,}
\end{equation}
where we introduced $\sigma_{\omega}$ as the wave packet's width in the frequency domain, which satisfies the relation $\sigma_{t}\cdot\sigma_{\omega}=1$.
From the pulse duration $T_{P} =  2 \sqrt{\ln{2}} \ \sigma_{t}$ the pulse spectral width $W_{P} =  2 \sqrt{\ln{2}} \ \sigma_{\omega} = \left( 2 \sqrt{\ln 2} \right)^{2} T_{P}^{-1}$ can be evaluated and be compared to the resonance width $\Gamma_{A}$. For example, if we consider the iron isotope ${}^{57}\text{Fe}$ with its small resonance width of $\hbar\cdot\Gamma_{A}=4.7$ neV driven by a free electron laser pulse of duration $T_{P}=100$ fs (FWHM),  the spectral width $W_{P}$ of the whole pulse is $10^6$ times larger than the resonance width $\Gamma_{0}$.
Based on Eq.~(\ref{Energy_EM}) we can calculate the energy per target area of a linear polarized x-ray pulse in the frequency domain by the use of Parseval's theorem 
\begin{equation}
\int_{-\infty}^{\infty} \left|E_{0}(t)\right|^{2} \mathrm{d}t = \int_{-\infty}^{\infty} |\hat{E}_{0}(\omega)|^{2} \mathrm{d}\omega \ \text{.}
\end{equation}
If the nuclear resonance width is small compared to the x-ray pulse bandwidth, then one can approximate the spectral overlap of pulse and resonance as 
\begin{equation}
c\epsilon_{0}\int_{-\frac{\Gamma_{A}}{2}}^{\frac{\Gamma_{A}}{2}} \hat{E}_{0}^{2}(\omega) \mathrm{d}\omega \approx c\epsilon_{0} \hat{E}_{0}^{2}(\omega_{0}) \Gamma_{A} \ \text{.}
\end{equation}
Finally, this expression has to be identical to the energy per target area
\begin{equation}
c\epsilon_{0}\Gamma_{A}\mathcal{E}^{2}_{0}\sigma_{t} \overset{!}{=} \frac{n_{res}\cdot\hbar\omega_{0}}{A_{target}} \ \text{,}
\end{equation}
which allows us to fix the constant
\begin{equation}
\mathcal{E}_{0}= \left(\frac{n_{res}\cdot\hbar\omega_{0}}{c A_{target}\epsilon_{0}\Gamma_{A}\sigma_{t}^{2}}\right)^{1/2} \ \text{.}
\end{equation}

\subsection{${}^{57}\text{Fe}$ nuclei in a cavity}

For our analysis of stimulated emission with x-rays interacting with ${}^{57}\text{Fe}$ M\"ossbauer nuclei, we consider a solid state target with a large ensemble of nuclei. In this case, it is not possible to determine the nucleus which participated in coherent scattering of the x-ray photon, such that the different possible scattering pathways interfere. If the scattering particles are placed in a suitable geometry~\cite{RoehlsbergerScience,Joerg12,PhysRevA.87.053837}, the intermediate scattering state can be well approximated as an exitonic state of the form~\cite{Dicke54,Hannon68,Hannon69,Kagan65,Kagan99,ScullyPRL06,ScullyPRL09,RoehlsbergerScience}
\begin{equation}\label{Exciton}
\ket{T_{\textbf{k}_{0}}}^{}=\frac{1}{\sqrt{N_{coh}}}\sum_{j} e^{i\textbf{k}_{0} \cdot \textbf{r}_{j} } \ket{e_{j},\textbf{0}} \ \text{,}
\end{equation}
where $\textbf{k}_{0}$ is the incident photon wave vector, $\textbf{r}_j$ is the position of nucleus $j$, and $N_{coh}$ is the number of excited nuclei within the coherence volume of interfering nuclei. As a result, the coupling of the x-rays to the nuclei is superradiantly enhanced.
Note that the preparation of the state Eq.~(\ref{Exciton}) and its subsequent superradiant exponential decay was recently experimentally demonstrated~\cite{RoehlsbergerScience}, enabled by placing the nuclei in a thin film cavity. 

For our simulations, we consider   parameters comparable to those of this experiment, but with variable number of resonant photons per pulse and pulse durations to simulate different potential light sources. The special cavity geometry allows us to reduce a $N_{coh}$-particle system to a simple two-level system with the collective state $\ket{T_{\textbf{k}_{0}}}$ as excited state~\cite{PhysRevA.88.043828}.

The decay rate for a single ${}^{57}\text{Fe}$ nucleus is $\Gamma_{A}=7.1\cdot10^{6} \ \text{s}^{-1}$ and the single dipole moment is $|d_{eg}|=1.3\cdot 10^{-35}$ Cm. To account the collective effects due to superradiance we replace the Rabi frequency by $\Omega_{N_{coh}}(t)=\sqrt{N_{coh}}\Omega(t)$ and the 
decay rate by  $\Gamma_{N_{coh}}=N_{coh}\Gamma_{A}$. Furthermore we use the 3D Wigner-Weisskopf decay rate
\begin{equation}
\Gamma_{A}=\frac{\omega_{A}^{3}|d_{eg}|^{2}}{3\pi\epsilon_{0} \hbar c^{3}} \ \text{.}
\end{equation}
The X-ray laser pulse with beam size $A_{beam}$ is irradiated on the target at a small angle $\varphi \approx 2.5\text{mrad}$, since then the x-ray light resonantly couples into one of the cavity modes~\cite{RoehlsbergerScience}. Thus the effective irradiated area on the target is enhanced to $A_{target}=d_{beam}^{2}/\sin\varphi$. The thickness of the ultra thin ${}^{57}\text{Fe}$ layer inside the cavity is $L_{target}=1.2\text{nm}$, but because of many reflections, a photon experiences an effective layer thickness of $L_{eff}\approx L_{target}\cdot Q$, with $Q<100$ in a planar low-$Q$ cavity. The $Q$ factor is proportional to the average lifetime of a resonant photon inside a cavity and therefore also proportional to the number of reflections inside a cavity. Thus the irradiated target volume is $V_{target}=A_{target}\cdot L_{eff}$ with the corresponding target mass 
$m_{target}=\varrho_{Fe}\cdot V_{target}$. From this it follows that the number of irradiated nuclei is $N_{n}=m_{target}/m_{Fe}$, where $m_{Fe}$ is the mass of a single ${}^{57}\text{Fe}$ nucleus. We consider an enhancement of the collective decay rate $\Gamma_{N_{coh}}=25 \Gamma_{A}$, such that there are $N_{coh}=25$ nuclei inside the coherence volume and the number of coherence volumes inside the irradiated volume is $M_{coh}=N_{n}/N_{coh}$. Multiplying the number of coherence volumes inside the irradiated volume $M_{coh}$ with the excited state population $\rho_{22}$, the number of excited nuclei can be estimated. Furthermore,  the event rate of stimulated emission is obtained via $M_{coh}\cdot N_{se}$.  Here, $N_{se}$ is the number of emitted photons due to stimulated emission, as introduced in Eq.~(\ref{Nse}).

Another aspect of the excitonic superposition Eq.~(\ref{Exciton}) is that the superradiant emission is highly directional in forward direction, where contructive interference beteen all possible scattering pathways occurs. Therefore, the standard experimental approach is to observe the scattered light in forward direction (nuclear forward scattering).

 \begin{figure}[t]
 \centering
    \includegraphics[width=0.8\columnwidth]{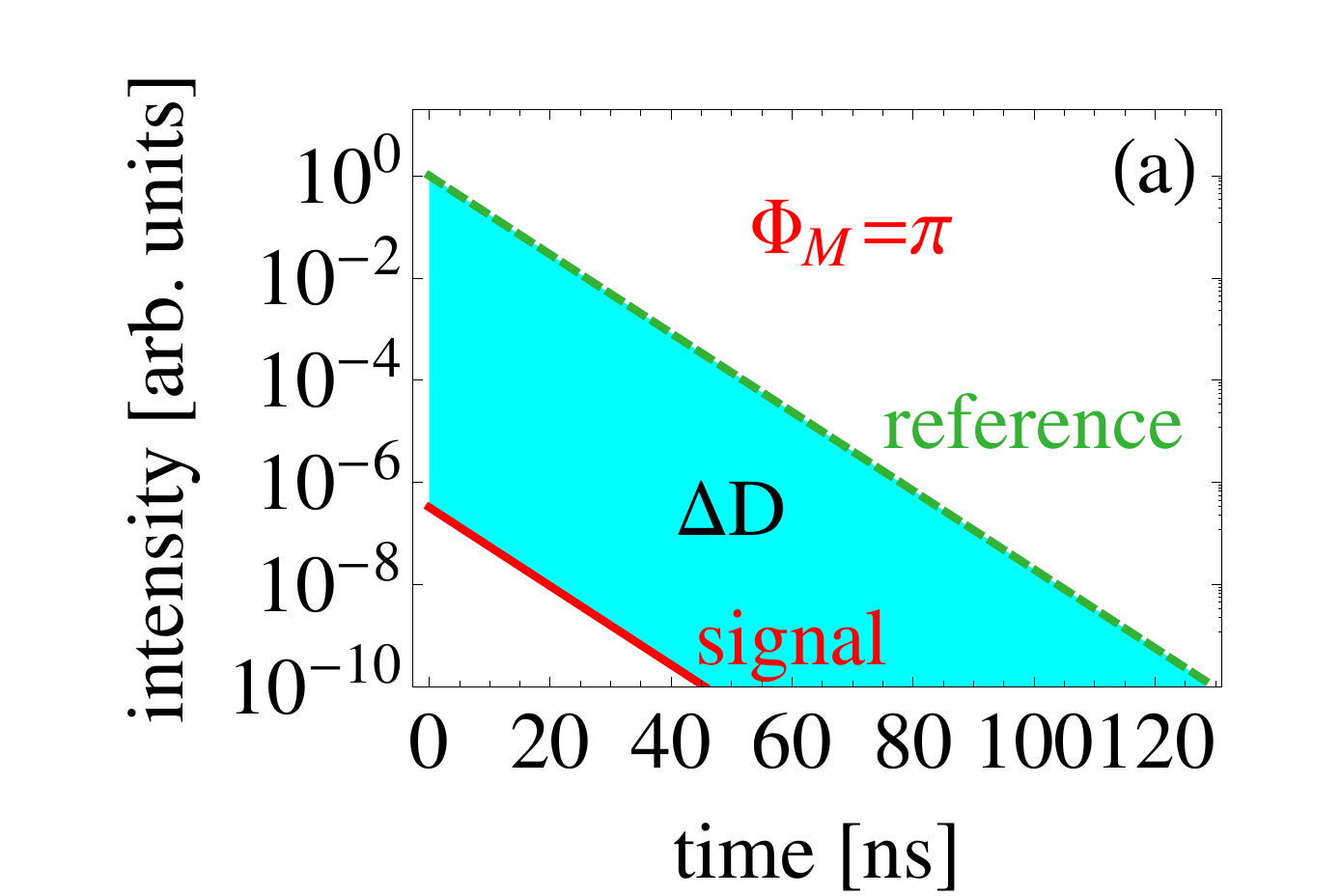}\\
    \includegraphics[width=0.8\columnwidth]{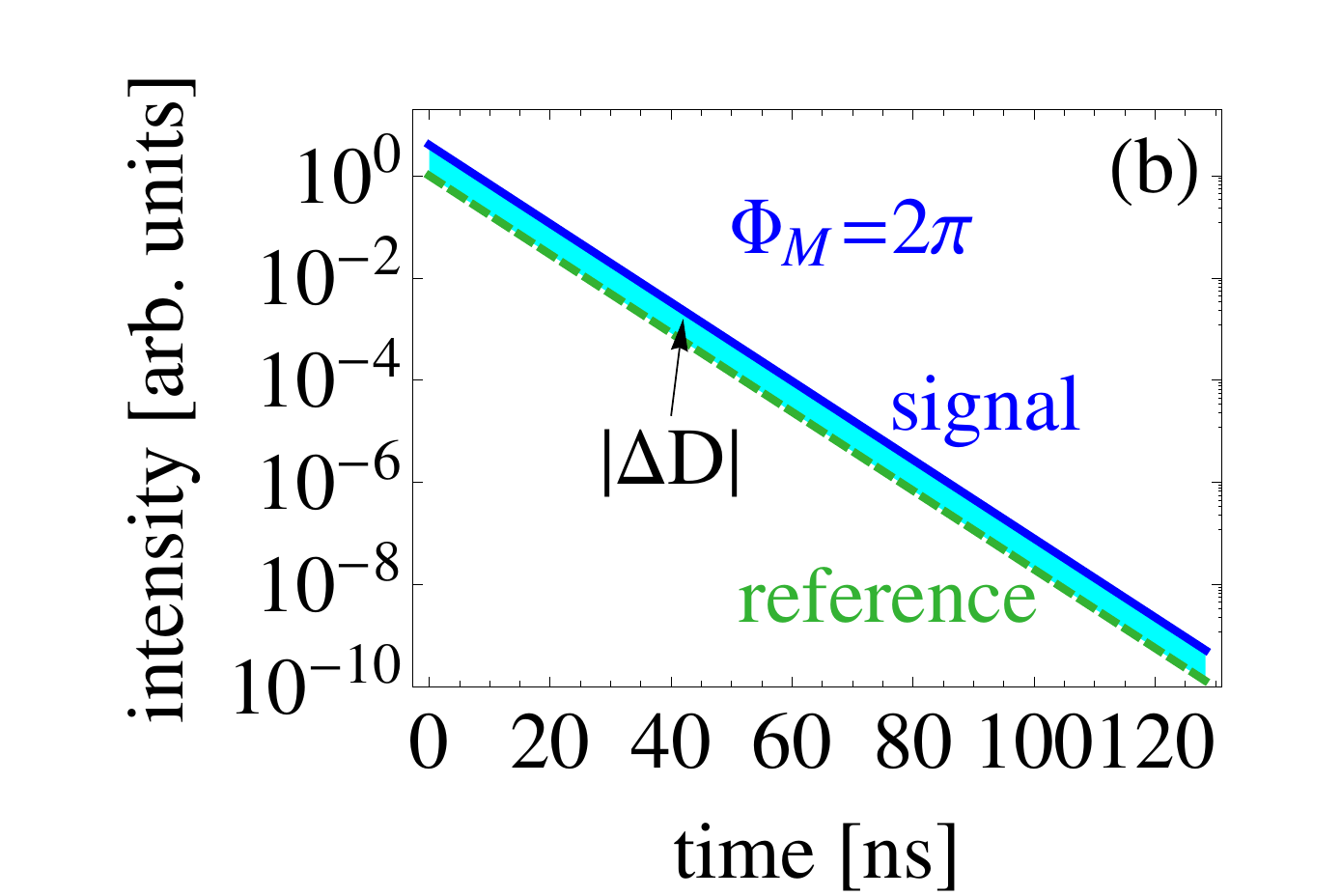}
    
\caption{\label{ResIntensity-fel}(Color online) Stimulated emission signatures for free electron laser parameters (pulse duration $T_{P}=100fs$ and bunch separation $\tau_{d}=5ps$) (a) for  stimulated emission ($\Phi_{M}=\pi$), (b) for  absorption ($\Phi_{M}=2\pi$). Both pulses  are assumed to have the same resonant intensity.  The dashed reference corresponds to spontaneous decay without a second pulse. The solid line shows results with a second pulse at $\tau_{d}$. The shaded area corresponds to $\Delta D$. Note that the maxima of both plots are normalized to unity and therefore independent of the number of resonant photons $n_{res}$. }
\end{figure} 

\subsection{Numerical results}

Because of the large number of non-resonant photons in X-ray pulses delivered by synchrotron or free electron laser sources it is challenging to directly detect the enhancement of the stimulating pulse due to stimulated emission. Therefore, we focus on the second indicator, i.e., the modification of the spontaneous decay following the stimulated emission. In view of typical nuclear forward scattering setups, we analyze this modification of the light scattered in forward direction.

We employ the pure spontaneous decay as reference, and scale the maximum of the time dependent intensity to unity in the subsequent analysis to enable better comparison between the different cases. We further define  for the case of two (exciting and stimulating) pulses 
\begin{eqnarray}
&&D_{signal}(\Phi_{M})=M_{coh}\cdot\frac{1}{(T_{end}-T_{start})}\nonumber \\
&&\int_{T_{start}}^{T_{end}}\rho_{22}(\Phi_{M},t=T_{start})e^{-\Gamma_{N_{coh}}t}\: \mathrm{d}t \ \text{,} 
\end{eqnarray}
which is a measure for the amount of light emitted within the interval from $T_{start}$ to $T_{end}$. 
$\Phi_{M}$ is the relative phase between the pulses. $T_{start}=\tau_{d}+T_{P}$ is chosen such that both pulses have already passed the nuclear target and the end time $T_{end}=-\ln(10^{-10})/\Gamma_{N_{coh}}$ is chosen such that the nuclear two-level system lost ten orders of magnitude of its population, covering most of the decay. Additionally we define a corresponding reference value with the single exciting pulse only as
\begin{eqnarray}
D_{ref}&=&M_{coh}\cdot\frac{1}{(T_{end}-T_{start})} \nonumber \\
&&\int_{T_{start}}^{T_{end}}  \rho_{22}(t=T_{start}) e^{-\Gamma_{N_{coh}}t}\: \mathrm{d}t  \,.
\end{eqnarray}
\begin{figure}[t]
 \centering
    \includegraphics[width=0.8\columnwidth]{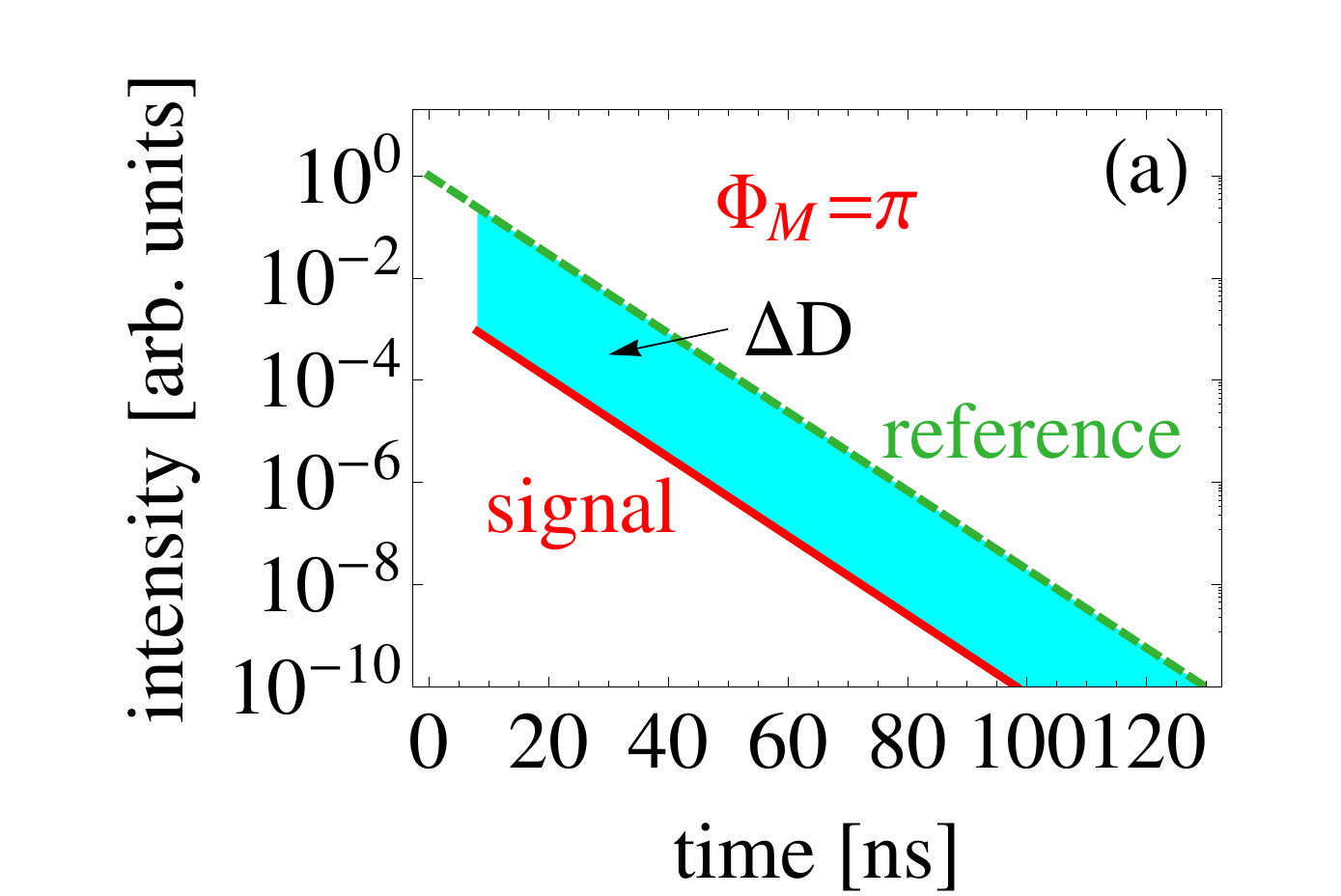}\\
    \includegraphics[width=0.8\columnwidth]{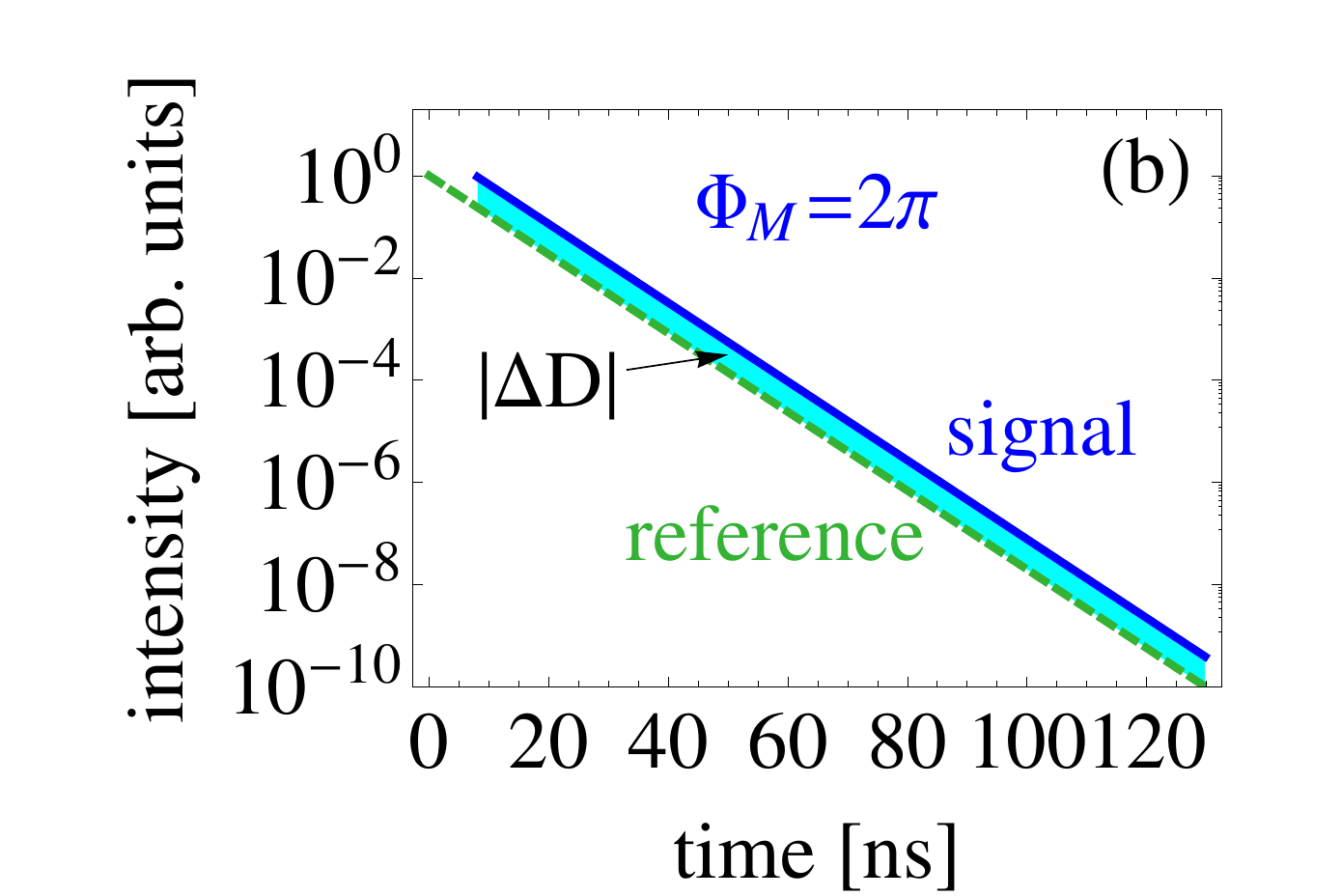}
     \caption{\label{ResIntensity}(Color online) Stimulated emission signatures for synchrotron-like parameters (pulse duration $T_{P}=100ps$ and bunch separation $\tau_{d}=8ns$) (a) for stimulated emission ($\Phi_{M}=\pi$), (b) for absorption ($\Phi_{M}=2\pi$). The dashed reference corresponds to spontaneous decay without a second pulse. The second pulse is assumed to have a number of resonant photons which is 1/4 of that of the first pulse.  The solid line shows results with a second pulse at $\tau_{d}$. The shaded area corresponds to $\Delta D$. Note that the maxima of both plots are normalized to unity and therefore independent of the number of resonant photons $n_{res}$. Note that in this figure, a fixed phase relation between the two light pulses is assumed, which is not the case at present synchrotron radiation sources.}
\end{figure} 
From these two quantities, we define the difference
\begin{equation}
\Delta D=D_{ref}-D_{signal}(\Phi_{M})
\end{equation} 
which is positive for stimulated emission and negative for absorption.

In Fig.~\ref{ResIntensity-fel}, results are shown for  pulse duration $T_{P}=100fs$ and pulse separation $\tau_{d}=5ps$. The pulse duration is a typical value for an x-ray free electron laser, and the pulse separation corresponds to values achievable in split-and-delay units (SDU) which are being developed for free electron lasers~\cite{XFEL}. We stress that the analysis presented here assumes a fixed phase relation between the two incident pulses, which could be achieved with a seeded free electron laser~\cite{seeding,seeding2} featuring long temporal coherence times, and by deriving the two pulses from a single initial pulse using the SDU. Note that the figure shows the results with maximum intensity scaled to unity. This way, the dependence on the mean number of photons in the incoming pulse is scaled out, and the different results can better be compared.  

Because of their short temporal separation, the nuclear target looses only a small fraction of its excitation in between the two pulses. Therefore, the absolute amount of population change induced by the second pulse is comparable to the initial excitation.  Depending on the relative phase between the pulses, the second pulse either increases the excitation of the nuclei via absorption, or reduces it via stimulated emission. 
We note that next to the relative phase, the effect of the second pulse also depends on its duration and intensity. The reason is that the nuclei essentially undergo fractional Rabi oscillations in the low-excitation regime, and the relative phase decides if those excite or de-excite the nuclei. Therefore, apart from the relative phase, the effect of the second pulse also can be controlled via the intensity of the second pulse, which could be manipulated, e.g., using absorbers in one arm of the SDU.

In Fig.~\ref{ResIntensity}, corresponding results are shown for pulse duration $T_P = 100$ps and bunch separation $\tau_d = 8$ns. The chosen pulse duration corresponds to a typical pulse duration at a synchrotron radiation source, and $\tau_d = 8$ns is the minimum separation between two electron bunches at the synchrotron radiation source PETRA III (DESY, Hamburg), as an example. We stress, however, that two pulses emitted from different electron bunches at synchrotron sources do not have a fixed phase relation relative to each other. Therefore the synchrotron results shown in Fig.~\ref{ResIntensity} could not be observed in a straightforward way. Also an SDU-like approach is challenging, since the longer pulse durations would require a longer delay time to separate the pulses, and since a splitting would only be meaningful with a temporally coherent pulse containing more than one resonant photon. Nevertheless, we show the results to visualize the effect of the very different time scales on the stimulated emission.  

It can be seen from Fig.~\ref{ResIntensity} that in the synchrotron case, due to the superradiant enhancement, most part of the excitation from the first pulse has already decayed before the second pulse arrives, even though the pulse separation is significantly shorter than the lifetime of a single nucleus. But while the absolute amount of population change due to the second pulse is low, on a logarithmic scale, it scan still clearly be identified (Note that it is possible to observe the nuclear decay induced by a single incident pulse over few orders of magnitude in the scattered light intensity~\cite{RoehlsbergerScience}). Again, the phase control between absorption and stimulated emission is clearly visible.

\section{Summary and discussion}

The aim of this work was to provide a detailed analysis of the spatio-temporal dynamics of stimulated emission, with the particular emphasis on stimulated emission with x-ray pulses interacting with nuclei. This setting is special, since typical modern x-ray sources deliver pulses with incident bandwidth exceeding the nuclear line width by several orders of magnitude. Since the non-resonant photons form an enormous background signal, it is not possible to observe the resonantly scattered light throughout the incident pulse duration. In conventional nuclear forward scattering, this problem is circumvented by imposing a time gating to observe the delayed light only, which is emitted after the incident pulse and thus the background signal has passed. If the nuclear lifetime is significantly longer than the incident pulse duration, still a large fraction of the scattered light due to the initial excitation can be observed this way.  

This immediately raises the question, whether stimulated emission can be observed in this delayed signal trailing the incident pulse as well. Interestingly, while stimulated emission is part of most related text books, the temporal dynamics of it is hardly studied at all. Therefore, to address the feasibility of the observation of nuclear stimulated emission, we first analyzed the general problem of the temporal dynamics of stimulated emission.

To this end, we studied an atom interacting with a quantized multimode electromagnetic field inside an ideal cavity. We found that the stimulated emission enhancement itself essentially overlaps  with the inducing pulse, which means that it could not be observed in the nuclear setting. But in addition to the enhancement, we identified a second signature, which is the {\it reduction} of the {\it delayed} scattered light intensity, arising from the accelerated de-excitation of the atoms throughout the stimulated emission. This second signature is well accessible in the nuclear setting. 

After having established these results, we switched to a semiclassical description, to facilitate the analysis for the nuclear case, which features several very different natural time scales which render a fully quantized treatment demanding. Here, next to the standard stimulated emission from an initially fully inverted target, we in particular studied the case of double pulse excitation in which the total excitation of the atoms remains low. If the two pulses are mutually coherent, then the effect of the second pulse strongly depends on the relative phase of the two pulses, which provides a convenient handle to detect stimulated emission. Finally, we analyzed stimulated emission in nuclei driven by free electron lasers or synchrotron radiation sources. Our analysis predicts that stimulated emission should be observable in the FEL case, if two pulses with fixed phase relation can be realized.  

Finally, we note that the results presented here depend on the model for the cooperative effects in the large ensemble of nuclei. On the one hand, these cooperative effects are associated to an enhancement of the spontaneous decay. On the other hand, they enhance the coupling to the driving x-ray field. This cooperative enhancement, however, depends on the size of the coherences volumes inside the medium, which are limited, for example, by photo absorption~\cite{Hannon_Trammell}. In this work, we chose the coherence volumes for spontaneous emission and for the x-ray excitation to be the same. If, however, the coherence volumes for the exciting pulse and for the radiative eigenmodes of the ensemble differ, then the temporal evolution of the ensemble and thus the amount of observable stimulated emission change, which we verified using numerical simulations with few atoms. To address this question, a full many-body analysis including propagational effects inside the medium would be required, which remains an open task for the future.

\bibliographystyle{myprsty}
\bibliography{references}{}

\begin{thebibliography}{10}

\bibitem{laser1}
P.~W. Milonni and J.~H. Eberly, {\em Laser Physics}, 2 ed. (Wiley \& Sons,
  Hoboken, 2010).

\bibitem{laser2}
A.~E. Siegman, {\em Lasers}, 1 ed. (University Science Books, Mill Valley, CA,
  1986).

\bibitem{ISI:000247344500011}
W. Ackermann {\it et~al.}, Nat. Phton. {\bf 1},  336  (2007).

\bibitem{ISI:000281467900020}
P. Emma {\it et~al.}, Nat. Photon. {\bf 4},  641  (2010).

\bibitem{ISI:000307046800014}
T. Ishikawa {\it et~al.}, Nat. Photon. {\bf 6},  540  (2012).

\bibitem{review-fel}
B.~W.~J. McNeil and N.~R. Thompson, Nat. Photon. {\bf 4},  814  (2010).

\bibitem{nina}
N. Rohringer, D. Ryan, R.~A. London, M. Purvis, F. Albert, J. Dunn, J.~D.
  Bozek, C. Bostedt, A. Graf, R. Hill, S.~P. Hau-Riege, and J.~J. Rocca, Nature
  {\bf 481},  488  (2012).

\bibitem{AnwendungKernlaser}
D.~E. Murnick and M.~S. Feld, Annual Review of Nuclear and Particle Science
  {\bf 29},  411  (1979).

\bibitem{SummaryNuklearLaser1}
G.~C. Baldwin and J.~C. Solem, Rev. Mod. Phys. {\bf 69},  1085  (1997).

\bibitem{SummaryNuklearLaser2}
L.~A. Rivlin, Quantum Electronics {\bf 37},  723  (2007).

\bibitem{ProposalNuclearLaser}
E.~V. Tkalya, Phys. Rev. Lett. {\bf 106},  162501  (2011).

\bibitem{BrinkeExciton}
N. ten Brinke, R. Sch\"utzhold, and D. Habs, Phys. Rev. A {\bf 87},  053814
  (2013).

\bibitem{BuzekSE}
G. Drobn\'y, M. Havukainen, and V. Bu\ifmmode~\check{z}\else \v{z}\fi{}ek,
  Journal of Modern Optics {\bf 47},  851  (2000).

\bibitem{Valente2}
D. Valente, Y. Li, J.~P. Poizat, J.~M. G\'erard, L.~C. Kwek, M.~F. Santos, and
  A. Auff\`eves, New Journal of Physics {\bf 14},  083029  (2012).

\bibitem{Shanhui}
E. Rephaeli and S. Fan, Phys. Rev. Lett. {\bf 108},  143602  (2012).

\bibitem{Elyutin2012}
P.~V. Elyutin, Phys. Rev. A {\bf 85},  033816  (2012).

\bibitem{Valente1}
D. Valente, S. Portolan, G. Nogues, J.~P. Poizat, M. Richard, J.~M. G\'erard,
  M.~F. Santos, and A. Auff\`eves, Phys. Rev. A {\bf 85},  023811  (2012).

\bibitem{RoehlsbergerScience}
R. R\"ohlsberger, K. Schlage, B. Sahoo, S. Couet, and R. R\"uffer, Science {\bf
  328},  1248  (2010).

\bibitem{sgc}
K.~P. Heeg, H.-C. Wille, K. Schlage, T. Guryeva, D. Schumacher, I. Uschmann,
  K.~S. Schulze, B. Marx, T. K\"ampfer, G.~G. Paulus, R. R\"ohlsberger, and J.
  Evers, Phys. Rev. Lett. {\bf 111},  073601  (2013).

\bibitem{RR12}
R. R\"ohlsberger, H.-C. Wille, K. Schlage, and B. Sahoo, Nature {\bf 482},  199
   (2012).

\bibitem{olga}
F. Vagizov, V. Antonov, Y.~V. Radeonychev, R.~N. Shakhmuratov, and O.
  Kocharovskaya, Nature {\bf 508},  80  (2014).

\bibitem{PRLEvers06}
T.~J. B\"urvenich, J. Evers, and C.~H. Keitel, Phys. Rev. Lett. {\bf 96},
  142501  (2006).

\bibitem{Adams13}
B.~W. Adams, C. Buth, S.~M. Cavaletto, J. Evers, Z. Harman, C.~H. Keitel, A.
  P\'alffy, A. Pic\'on, R. R\"ohlsberger, Y. Rostovtsev, and K. Tamasaku,
  Journal of Modern Optics {\bf 60},  2  (2013).

\bibitem{EversPRL09}
A. P\'alffy, C.~H. Keitel, and J. Evers, Phys. Rev. Lett. {\bf 103},  017401
  (2009).

\bibitem{PhysRevLett.109.197403}
W.-T. Liao, A. P\'alffy, and C.~H. Keitel, Phys. Rev. Lett. {\bf 109},  197403
  (2012).

\bibitem{PhysRevLett.109.262502}
W.-T. Liao, S. Das, C.~H. Keitel, and A. P\'alffy, Phys. Rev. Lett. {\bf 109},
  262502  (2012).

\bibitem{Sturhahn}
W. Sturhahn, Journal of Physics: Condensed Matter {\bf 16},  S497  (2004).

\bibitem{BuzekKim1997}
V. Bu\ifmmode~\check{z}\else \v{z}\fi{}ek and M.~G. Kim, Journal of the Korean
  Physical Society {\bf 30},  413  (1997).

\bibitem{BuzekKnight}
V. Bu\ifmmode~\check{z}\else \v{z}\fi{}ek, G. Drobn\'y, M.~G. Kim, M.
  Havukainen, and P.~L. Knight, Phys. Rev. A {\bf 60},  582  (1999).

\bibitem{Buzek2000}
M. Havukainen, G. Drobn\'y, S. Stenholm, and V. Bu\ifmmode~\check{z}\else
  \v{z}\fi{}ek, Journal of Modern Optics {\bf 46},  1343  (1999).

\bibitem{Scully}
M.~O. Scully and M.~S. Zubairy, {\em Quantum optics}, 6. ed. (Cambridge Univ.
  Press, Cambridge, 2008).

\bibitem{Moessbauer}
R.~L. M\"ossbauer, Zeitschrift f\"ur Physik {\bf 151},  124  (1958).

\bibitem{Jackson}
J.~D. Jackson, {\em Classical electrodynamics}, 3. ed. (Wiley, New York, 1999).

\bibitem{Ficek}
Z. Ficek and S. Swain, {\em Quantum interference and coherence} (Springer, New
  York, 2005).

\bibitem{Kiffner_Vacuum_Processes}
M. Kiffner, M. Macovei, J. Evers, and C. Keitel, {\em Progress in Optics}
  (Elsevier Science, Burlington, 2010), Vol.~55, pp.\ 85--197.

\bibitem{LindbladSuper}
G. Lindblad, Communications in Mathematical Physics {\bf 48},  119  (1976).

\bibitem{Joerg12}
Y. Li, J. Evers, H. Zheng, and S.-Y. Zhu, Phys. Rev. A {\bf 85},  053830
  (2012).

\bibitem{PhysRevA.87.053837}
Y. Li, J. Evers, W. Feng, and S.-Y. Zhu, Phys. Rev. A {\bf 87},  053837
  (2013).

\bibitem{Dicke54}
R.~H. Dicke, Phys. Rev. {\bf 93},  99  (1954).

\bibitem{Hannon68}
J.~P. Hannon and G.~T. Trammell, Phys. Rev. {\bf 169},  315  (1968).

\bibitem{Hannon69}
J.~P. Hannon and G.~T. Trammell, Phys. Rev. {\bf 186},  306  (1969).

\bibitem{Kagan65}
A.~M. Afanas'ev and Y. Kagan, Sov. Phys. JETP {\bf 21},  215  (1965).

\bibitem{Kagan99}
Y. Kagan, Hyperfine Interactions {\bf 123-124},  83  (1999).

\bibitem{ScullyPRL06}
M.~O. Scully, E.~S. Fry, C.~H.~R. Ooi, and K. W\'odkiewicz, Phys. Rev. Lett.
  {\bf 96},  010501  (2006).

\bibitem{ScullyPRL09}
M.~O. Scully, Phys. Rev. Lett. {\bf 102},  143601  (2009).

\bibitem{PhysRevA.88.043828}
K.~P. Heeg and J. Evers, Phys. Rev. A {\bf 88},  043828  (2013).

\bibitem{XFEL}
S. Roling, S. Braun, P. Gawlitza, L. Samoylova, B. Siemer, H. Sinn, F. Siewert,
  F. Wahlert, M. W\"ostmann, and H. Zacharias, Proc. SPIE {\bf 8778},
  (2013).

\bibitem{seeding}
J. Amann {\it et~al.}, Nature Photon. {\bf 6},  693  (2012).

\bibitem{seeding2}
E. Allaria {\it et~al.}, Nature Photon. {\bf 7},  913  (2013).

\bibitem{Hannon_Trammell}
J. Hannon and G. Trammell, Hyperf. Int. {\bf 123-124},  127  (1999).

\end{thebibliography}

\end{document}